\documentclass[usenatbib]{mnras}

\usepackage{newtxtext,newtxmath}

\usepackage[T1]{fontenc}

\DeclareRobustCommand{\VAN}[3]{#2}
\let\VANthebibliography\thebibliography
\def\thebibliography{\DeclareRobustCommand{\VAN}[3]{##3}\VANthebibliography}

\usepackage{graphicx}	
\usepackage{amsmath}	

\usepackage{amssymb}	
\usepackage{cite}
\usepackage{hyperref}
\usepackage{color}
\usepackage{bm}

\newcommand{\sgra}{Sgr A$^{*}$}
\definecolor{darkgreen}{RGB}{0,100,0}

\defcitealias{Li2017}{L17}
\defcitealias{Yuan2009}{Y09}
\defcitealias{GRAVITY2018}{G18}
\defcitealias{GRAVITY2023}{G23}

\title[\sgra\ flares]{Revisiting flares in Sagittarius A$^*$ based on general relativistic magnetohydrodynamic numerical simulations of black hole accretion }

\author[Lin \& Yuan]{
Xi Lin$^{1,2}$,
and Feng Yuan$^{3,1,2}$\thanks{E-mail:fyuan@fudan.edu.cn}
\\
$^{1}$Shanghai Astronomical Observatory, Chinese Academy of Sciences, Shanghai 200030, People’s Republic of China\\
$^{2}$University of Chinese Academy of Sciences, 19A Yuquan Road, Beijing 100049, People’s Republic of China\\
$^{3}$Center for Astronomy and Astrophysics and Department of Physics, Fudan University, Shanghai 200438, People’s Republic of China}

\date{Accepted XXX. Received YYY; in original form ZZZ}
\pubyear{2022}
\begin{document}
\label{firstpage}
\pagerange{\pageref{firstpage}--\pageref{lastpage}}
\maketitle

\begin{abstract}
High-resolution observations with GRAVITY-VLTI instrument have provided abundant information about the flares in Sgr A*, the supermassive black hole in our Galactic center, including the time-dependent location of the centroid (a ``hot spot''), the light curve, and polarization.  Yuan et al. (2009) proposed a ``coronal mass ejection'' model to explain the  flares and their association with the plasma ejection. The key idea is that magnetic reconnection in the accretion flow produces the flares and results in the formation and ejection of flux ropes. 
The dynamical process proposed in the model has been confirmed by three-dimensional GRMHD simulations in a later work. Based on this scenario, in our previous works the radiation of the flux rope has been calculated analytically and compared to the observations. In the present paper, we develop the model by directly using numerical simulation data to interpret observations. We first identify flux ropes formed due to reconnection from the data. By assuming that electrons are accelerated in the reconnection current sheet and flow into the flux rope and emit their radiation there, we have calculated the time-dependent energy distribution of electrons after phenomenologically considering their injection due to reconnection acceleration, radiative and adiabatic cooling. The radiation of these electrons is calculated using the ray-tracing approach. The trajectory of the hot spot, the radiation light curve during the flare, and the polarization are calculated. These results are compared with the GRAVITY observations and good consistencies are found. 

\end{abstract}


\section{Introduction}

As the closest super-massive black hole (SMBH; mass $M = 4.297\times10^{6} \ {M_\mathrm{\odot}}$, distance $D=8.277\mathrm{kpc}$, \citealt{GRAVITY2022}), Sagittarius A* (Sgr A*), which resides in the Galactic center, provides an ideal laboratory for studying accretion physics in the extreme gravitational field. One of the most fascinating observational results is the  flares in near-infrared (NIR, \citealt{Genzel2003,Ghez2004,Dodds-Eden2011,Do2019,Haggard2019,Witzel2021}) and X-ray \citep{Baganoff2001,Baganoff2003,Neilsen2013,Ponti2015} wavebands. The typical timescale of the NIR flare is about 30 minutes to 1 hour with a frequency of $1-4$ times a day. During the flares, the flux can be enhanced by more than 10 times. An X-ray flare is often observed to accompany the NIR flare (but not vice versa, \citealt{Zadeh2012,Fazio2018}), suggesting they may share the same physical origin. The observed NIR polarization degree of $20-40\%$ suggests that the NIR radiation most likely originates from synchrotron emission of energetic electrons \citep{Yuan2004,Shahzamanian2015,GRAVITY2018}. The X-ray flare may be powered in the same process from electrons with higher energy or originate from Comptonized synchrotron photons \citep{Yuan2004,Dodds-Eden2010,Zadeh2012}. Besides the flare state, Sgr A* spends most of its time in the quiescent state characterized by continuous and low-amplitude flux variability. In the quiescent state, its bolometric luminosity is as low as $L_\mathrm{bol} \sim 6.8-9.2 \times 10^{35} \ \mathrm{erg}\ \mathrm{s}^{-1} \sim 10^{-9} \ L_\mathrm{Edd} $, where ${L_\mathrm{Edd}}$ is the Eddington luminosity \citep{Bower2019,EHT2022a}. It has been well established that such a low luminosity arises from the radiatively inefficient accretion flow (RIAF) around the black hole \citep{1994ApJ...428L..13N,1995ApJ...452..710N,Yuan2003,YuanNarayan2014,EHT2022e}. 

Recently, Sgr A* has also been one of the primary targets for VLTI-GRAVITY \citep{GRAVITY2018,GRAVITY2020orb,GRAVITY2020pol,GRAVITY2023}. The high precision observation makes it possible to resolve Sgr A* both spatially (with an astrometric accuracy of $\sim 10 \mathrm{\upmu as} \approx 2r_{\mathrm{g}}$, where $r_{\mathrm{g}}=GM/c^2)$ and temporally. During the flares, NIR flux centroid is found to exhibit a clockwise continuous rotation with a projected radius of $\sim 10r_{\mathrm{g}}$ from the central black hole and the rotation period is compatible with the duration of the flares. Besides, the polarization angle also rotates smoothly with a similar period. The observations can be well explained by a “hot spot” of infrared synchrotron emission rotating at $\sim 6-10r_\mathrm{g}$ from the black hole in the environment of a strong poloidal magnetic field \citep{Broderick2006,GRAVITY2018,GRAVITY2020orb}.
One of the most important constraints comes from the July 22 flare, which has the best data quality among the three flares observed in \citet{GRAVITY2018}. As is first pointed out in \citet{Matsumoto2020}, this flare shows a super-Keplerian motion of the brightness centroid. This rotation speed is higher than that of the accretion flow since the rotation of the accretion flow should be sub-Keplerian \citep{YuanNarayan2014}. A number of models have been proposed to relieve the tension \citep{Matsumoto2020,Ball2021,Lin2023,Aimar2023}. 

To understand the physical mechanism of flares, many  general-relativistic magnetohydrodynamic (GRMHD) and particle-in-cell simulations have been performed \citep{Nathanail2020,Ripperda2020,Dexter2020,Porth2021,Chatterjee2021,Mellah2022,Mellah2023,Jiang2023}. Among all the models, magnetic reconnection is now believed to be the most promising mechanism to accelerate electrons and produce flares. Current sheets and plasmoids have been found in many  simulations serving as the evidence for reconnection \citep{Nathanail2022,Ball2018a,Petersen2020,Ripperda2021a,Miki2022}. However, few of these works put effort into the interpretation of GRAVITY observational results.

\begin{figure}	\includegraphics[width=\columnwidth]{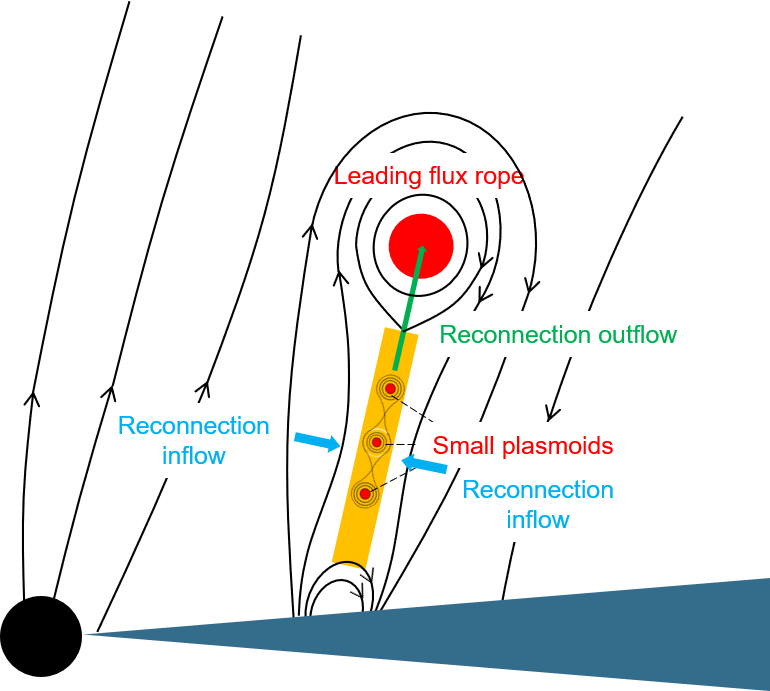}
    \caption{A schematic illustration of our model. Between the open field lines anchored on the black hole and mid-plane accretion flow, a magnetic arcade structure emerges out. Magnetic reconnection is triggered when the lines with opposite polarity approached close enough. Small plasmoids are formed in the current sheet due to tearing mode which are eventually advected outward and converge at the leading flux rope.}
    \label{fig:sketch}
\end{figure}

In this work, we make use of GRMHD simulation data to explain the GRAVITY results, including light curve, centroid motion and polarization rotation. This is the fifth paper in our series of works to investigate the flares of Sgr A* based on the model of ``reconnection-driven ejected flux rope''. This model was first proposed in \citet{Yuan2009}. 
The basic picture of the model is demonstrated in Figure \ref{fig:sketch}. A magnetic arcade emerges from the accretion disk due to Parker instability \citep{Parker1966}. Due to some physical processes, such as the differential rotation and turbulent motion of the accretion flow, the field lines with different polarities could come close enough. In this case, a current sheet will be formed and magnetic reconnection will be triggered. Some magnetic field lines will become closed and some plasma will be locked into the closed field lines and form a flux rope. The reconnection will accelerate electrons and the radiation of these energetic electrons will be responsible for the flares. The flux ropes, once they are formed, will be ejected out by magnetic forces.
A prediction of the model is therefore the physical association of the flares and  plasma ejection. It should be noted that the flux ropes in our work are not equivalent to the plasmoids formed in the current sheet due to tearing mode instability as proposed in some works. As is shown in Figure \ref{fig:sketch},  a flux rope is a field topology characterized by bundles of helical magnetic field lines collectively spiraling around a common axis, and these field lines originate from the reconnected magnetic arcades.

This prediction seems to have some observational evidence \citep{Yuan2009}. In many black hole sources, flares are observed to be associated with the blob ejection. Examples include radio galaxy 3C120 \citep{Chatterjee2009,Casadio2015},  blazar PKS 1510–089 \citep{Park2019}, black hole X-ray binary GRS 1915+105 \citep{Mirabel1994, Fender1999, Jones2005}, low-luminosity AGN M81 \citep{King2016} and M87 \citep{Hada2014}. The time delay between lower frequency radio flares and higher frequency radio flares in Sgr A* is another strong evidence for the ejection and expansion of the ejected discrete plasma blob \citep{2006ApJ...650..189Y,Dodds-Eden2009,Dodds-Eden2010}. 

\citet{Li2017} applied the scenario proposed in \citet{Yuan2009} to interpret the flares observed in Sgr A*. They have calculated the dynamics of the flux rope after its ejection and the time-dependent energy distribution of accelerated  electrons and their radiation. The model successfully explains the NIR/X-ray light curves as well as the spectrum during Sgr A* flares. This  work was recently improved by \citet{Lin2023} in the following two aspects, aiming at understanding the new observational results by GRAVITY. One is that, the toroidal motion of the ejected flux rope, which was neglected for simplicity in the previous work, was taken into account. The second improvement is that, the radiative transfer of the accelerated electrons was precisely calculated using the ray-tracing approach, which is to consider the important effects of general relativity close to the black hole.  The model was compared to the GRAVITY results and has well explained the observed light curve of the flares, the time-dependent distance of the hot spot, and why the light curve of some flares has double peaks. The super-Keplerian motion of the hot spot has also been discussed.   

All the above-mentioned three works are analytical. It is thus crucial to examine the model by detailed MHD numerical simulations. To examine the basic dynamical scenario proposed in \citet{Yuan2009}, \citet{Miki2022} have performed 3D GRMHD numerical simulations of accretion flows. They found that, as predicted by \citet{Yuan2009},  flux ropes are formed due to magnetic reconnection, which is likely driven by the turbulent motion and differential rotation of the accretion flow\footnote{It is worth noting that \citet{Nathanail2020} obtained similar results.}. They further found that flux ropes formed inside of 10–15$r_g$ mainly stay within the accretion flow, while flux ropes formed beyond this radius are ejected outward. Moreover,  all of the above-mentioned processes are found to occur quasi-periodically, with the period $\sim 1000 r_g/c$. 

In the present work, we develop the analytical \citet{Lin2023} work by directly using GRMHD simulation data of black hole accretion flows to explain the GRAVITY observations. Compared to \citet{Lin2023}, we  analyze the simulation data to find the flux rope and follow its trajectory, instead of solving an analytical dynamical equation. The magnetic field strength and configuration are also more realistic in the sense that they are directly taken from the simulation data rather than using a simplified analytical model. The calculation of the particle acceleration by reconnection and energy and spatial distribution is also based on the simulation data. The structure of the paper is as follows. In Section \ref{sec:MHDsimulation}, we introduce our GRMHD simulations of hot accretion flows. The distribution of current sheets and the formation and motion of flux ropes are investigated in sections \ref{rcs} and \ref{fluxrope}, respectively. In Section \ref{sec:nonthermalelectrons}, by solving a time-dependent evolution equation, we obtain the time-dependent non-thermal electron distribution of nonthermal electrons accelerated by magnetic reconnection based on the simulation data. The model predictions and and comparison with GRAVITY observations are given in Section \ref{sec:results}. We  summarize our results in Section \ref{sec:discussion}.

\section{Model}

\subsection{GRMHD numerical simulation}
\label{sec:MHDsimulation}
It is well established now that the accretion flow in Sgr A* is a radiatively inefficient hot accretion flow \citep{YuanNarayan2014}. There are two modes of hot accretion, namely SANE (standard and normal evolution) and MAD (magnetically arrested disk) \citep{Narayan2003,Igumenshchev2003,Sasha2011}. It is still uncertain which mode the accretion flow in Sgr A* is.  The polarization signature of Sgr A* prefers a vertical dominant magnetic field near the black hole \citep{Wielgus2022,GRAVITY2023,GRAVITY2020c}, which indicates the accretion disk of Sgr A* seems to be magnetically arrested (MAD, \citealt{Narayan2003,Sasha2011}).  Observations reveal around 30 massive Wolf–Rayet (WR) magnetized stars locating in the central parsec of Sgr A*'s rotation disk \citep{Paumard2006,Martins2007}. The accretion flow can be magnetically arrested by the continuous accretion of the magnetic field from large radii via the stellar winds of these stars \citep{Ressler2020}. However it is still confusing why Sgr A* is lack of a jet, which is a characteristic feature of MADs. The spin value of the black hole in Sgr A* is also not well determined.

Given this situation, to compare the difference between different accretion modes, we simulate three models, namely one SANE (standard and normal evolution) model with $a=0.98$ and two MAD models with $a=0$ and $a=0.98$. The simulation approach is described in detail in \citet{Yang2021} and \citet{Yang2024}. We briefly summarize some important features here for completeness. We employ {\tt ATEHA++} \citep{Athena2016,Athena2020} to solve general relativistic magnetohydrodynamics (GRMHD) equations in the Kerr metric, and the Kerr-Schild coordinates are adopted. The GRMHD equations describing the evolution of the
accretion flow are written as
\begin{center}
\begin{align}
&\nabla_\mu\left(\rho u^\mu\right)=0, \\
&\nabla_\mu T^{\mu \nu}=0, \\
&\nabla_\mu{ }^* F^{\mu \nu}=0,
\end{align}
\end{center}
with
\begin{center}
\begin{align}
& \ T^{\mu \nu}=\left(\rho h+b_\lambda b^\lambda\right) u^\mu u^\nu+\left(p+\frac{1}{2} b_\lambda b^\lambda\right) g^{\mu \nu}-b^\mu b^\nu, \\
&*F^{\mu \nu}=b^\mu u^\nu-b^\nu u^\mu.
\end{align}
\end{center}
Here $\rho$ is the rest-mass density, $p$ is the fluid pressure, $u^{\mu}$ is the fluid four-velocity, $T^{\mu \nu}$ is the energy-momentum tensor, $^* F^{\mu \nu}$ is the dual of the Faraday tensor, $b^{\mu}$ is the magnetic four-vector in the fluid comoving frame and $h(\rho,p)$ is the specific enthalpy of the fluid. The accretion starts from an axisymmetric hydrostatic equilibrium torus \citep{Fishbone1976} rotating around the black hole. The intial torus is threaded by a poloidal magnetic field with a different configuration for different models. For the MAD models, a single poloidal loop of magnetic field is set and enough magnetic flux is added inside the torus so that the flux can quickly be accumulated and reach a saturation limit via accretion. For the SANE, we initialize the magnetized torus with multiple poloidal magnetic loops of alternating polarity. The magnetic field lines with alternating polarity tend to reconnect when coming close enough, which will consume the magnetic flux and maintain the accretion flow in the SANE state. The code utilizes static mesh refinement (SMR) techniques to improve resolution in the areas of interest while maintaining the performance and applies the staggered-mesh constrained transport (CT) method to preserving a divergence-free magnetic field \citep{Evans1988}. 

All the simulation runs are performed up to $t=40000$. They correspond to 8.9 orbital periods of the accretion flow at the pressure maximum. The “inflow equilibrium’’ is reached about 80$r_g$ for the two MAD models and 30$r_g$ for the SANE model. The root grid resolution in the $r,\theta,\phi$ direction is $88 \times 32 \times 16$ for MAD98 and $110 \times 32 \times 16$ for MAD00 and SANE98. Different refinement criteria is adopted for different models and is summarized in Table \ref{tab:1} and \ref{tab:2}. The effective resolution $N_{r} \times N_\mathrm{{\theta}} \times N_\mathrm{{\phi}}$ for the three models and other model parameters describing the initial torus are listed in Table \ref{tab:3}. We note that the resolution of the present paper is at least two times higher in each of the three dimension than \citet{Yang2021}. The higher resolution is helpful for us to find flux ropes and obtain more reliable results. And it is also necessary as we will trace the flux rope over time. We have compared the main results obtained by this high resolution data with the low resolution data and found that the results are convergent. Readers can refer to Appendix \ref{appendix:A} for more details. 

\begin{table}
\centering
\caption{The SMR of MAD98. The root resolution is $88 \times 32 \times 16$.}
\begin{tabular} {lccc}
\hline \hline Level & $r / r_g$ & $\theta / \pi$ & $\phi / \pi$ \\
\hline 0 & {$[1.1,1200]$} & {$[0,1]$} & {$[0,2]$} \\
\hline 1 & {$[1.1,200]$} & {$[0.1305,0.8695]$} & {$[0,2]$} \\
\hline & {$[1.1,30]$} & {$[0.1942,0.8058]$} & {$[0,2]$} \\
2 & {$[6.4,1200]$} & {$[0,0.1210]$} & {$[0,2]$} \\
& {$[6.4,1200]$} & {$[0.8790,1]$} & {$[0,2]$} \\
\hline 3 & {$[36.7,1200]$} & {$[0,0.05730]$} & {$[0,2]$} \\
& {$[36.7,1200]$} & {$[0.9427,1]$} & {$[0,2]$} \\

\hline
\label{tab:1}
\end{tabular}
\end{table}

\begin{table}
\centering
\caption{The SMR of MAD00 and SANE98. The root resolution is $110 \times 32 \times 16$. }
\begin{tabular} {lccc}
\hline \hline Level & $r / r_g$ & $\theta / \pi$ & $\phi / \pi$ \\
\hline 0 & {$[1,1200]$} & {$[0,1]$} & {$[0,2]$} \\

\hline & {$[1,200]$} & {$[0.1305,0.8695]$} & {$[0,2]$} \\
1 & {$[1,30]$} & {$[0,0.1942]$} & {$[0,2]$} \\
& {$[1,30]$} & {$[0.8058,1]$} & {$[0,2]$} \\

\hline & {$[1,30]$} & {$[0.1942,0.8058]$} & {$[0,2]$} \\
2 & {$[6.4,1200]$} & {$[0,0.1210]$} & {$[0,2]$} \\
& {$[6.4,1200]$} & {$[0.8790,1]$} & {$[0,2]$} \\
\hline 3 & {$[36.7,1200]$} & {$[0,0.0573]$} & {$[0,2]$} \\
 & {$[36.7,1200]$} & {$[0.9427,1]$} & {$[0,2]$} \\
\hline
\label{tab:2}
\end{tabular}
\end{table}

\begin{table*}
\renewcommand\arraystretch{1.5}
 \caption{Setup for the various models. The first two columns are the black hole dimensionless spin and effective resolution. The last four columns are the parameters that determine the initial torus. $N$ and $\lambda_{B}$ set the number and the characteristic length scale of the poloidal loops. $r_\mathrm{start}$ and $r_\mathrm{end}$ denote the inner and outer edges of the magnetized region.}
 \setlength{\tabcolsep}{5mm}{
 \begin{tabular}{cccccccc}
 \hline \hline Model & $a$ & $N_r \times N_{\theta} \times N_{\phi}$ & $N$ & $\lambda_B$ &$r_\mathrm{start} (r_\mathrm{g}) $
 &$r_\mathrm{end} (r_\mathrm{g})$\\
 
  \hline SANE98 & 0.98 &  $880 \times 256 \times 128$ & 8 & 3.75 &25 &550\\
  MAD00 &0 & $880 \times 256 \times 128$ & 1 & 25 &25 &810 \\
  MAD98 &0.98 & $704 \times 256 \times 128$ & 1 & 25 &25 &810\\
 \hline
\label{tab:3}
\end{tabular}}
\end{table*}

\begin{figure}	\includegraphics[width=\columnwidth]{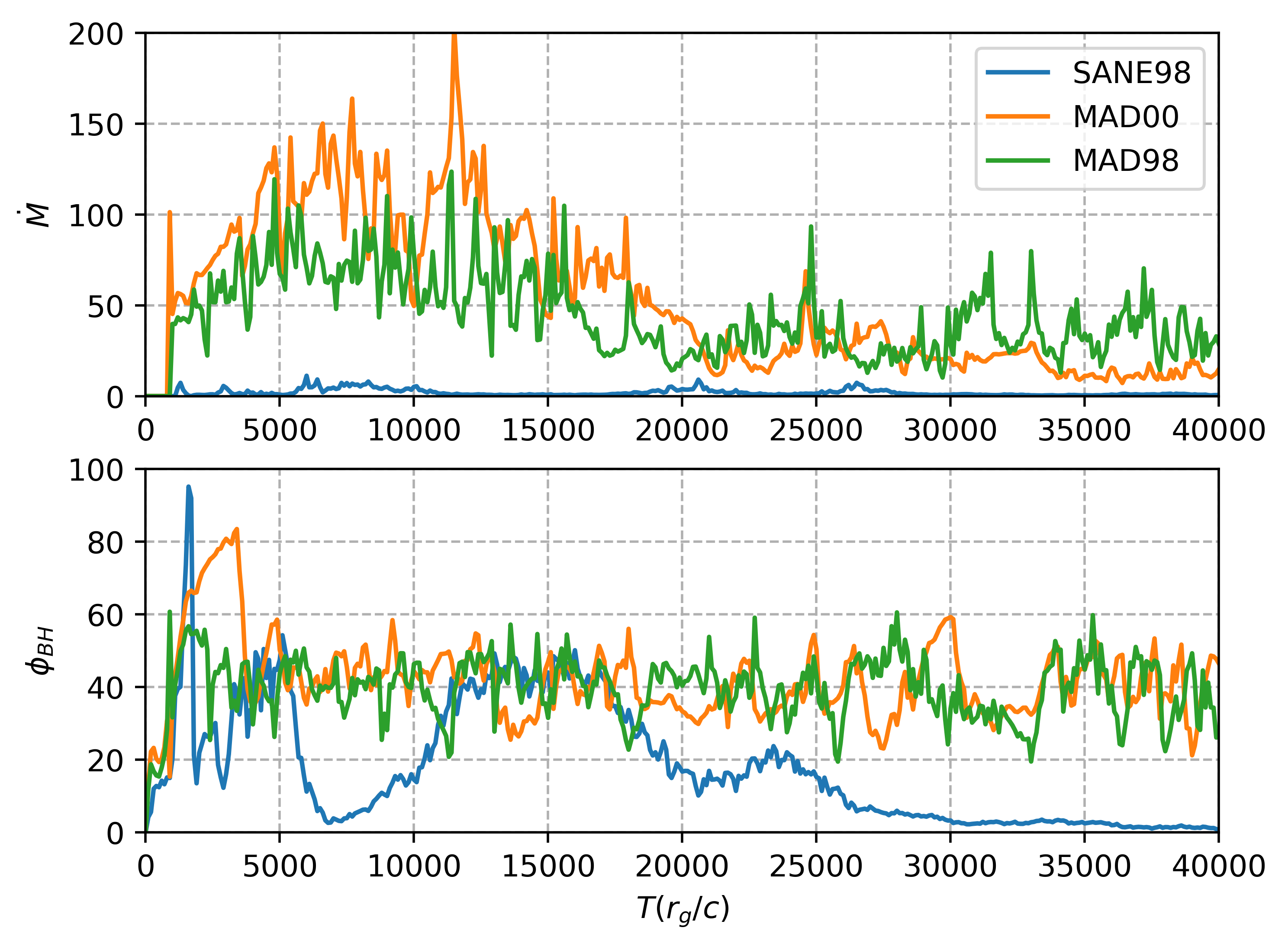}
    \caption{Mass accretion rate and normalized magnetic flux as functions of time.}
    \label{fig:mdot}
\end{figure}

To compare the difference between various accretion modes and investigate the influence of the black hole spin, we first calculate the accretion rate and magnetic flux brought into the black hole for each model. The mass accretion rate is calculated by
\begin{equation}
\dot{M}=-\int_0^{2 \pi} \int_0^\pi \rho u^r \sqrt{-g} \mathrm{~d} \theta \mathrm{d} \phi,
\end{equation}
where $g$ is the determinant of the metric. The integration is over all $\theta,\phi$ at the black hole horizon, which is defined by $r_{\mathrm{h}}=(1+\sqrt{1-a^2}) r_{\mathrm{g}}$. As is shown in the upper panel of Figure \ref{fig:mdot}, the accretion rate for all models has reached a roughly convergent value after $t\sim20000$, which indicates the accretion maintains at a quasi-stationary state. The difference between the two accretion modes can be reflected more clearly on the magnetic flux brought across the event horizon, which is defined by
\begin{equation}
\Phi_{\mathrm{BH}}=\frac{1}{2} \int_0^{2 \pi} \int_0^\pi \sqrt{4\pi}\left|B^r\right| \sqrt{-g} \mathrm{~d} \theta \mathrm{d} \phi.
\end{equation}
We show the normalized magnetic flux $\phi_{\mathrm{BH}}=\Phi_{\mathrm{BH}}/\sqrt{\dot{M}}$ at the lower panel of Figure \ref{fig:mdot}. The magnetic flux of  MADs fluctuates around $\phi_{\mathrm{BH}}\sim 40$ after $t=5000$, which is close to the saturation limit of MAD of $\phi_{\mathrm{max}}\sim 50$ \citep{Sasha2011}. 

\begin{figure}
	\includegraphics[width=\columnwidth]{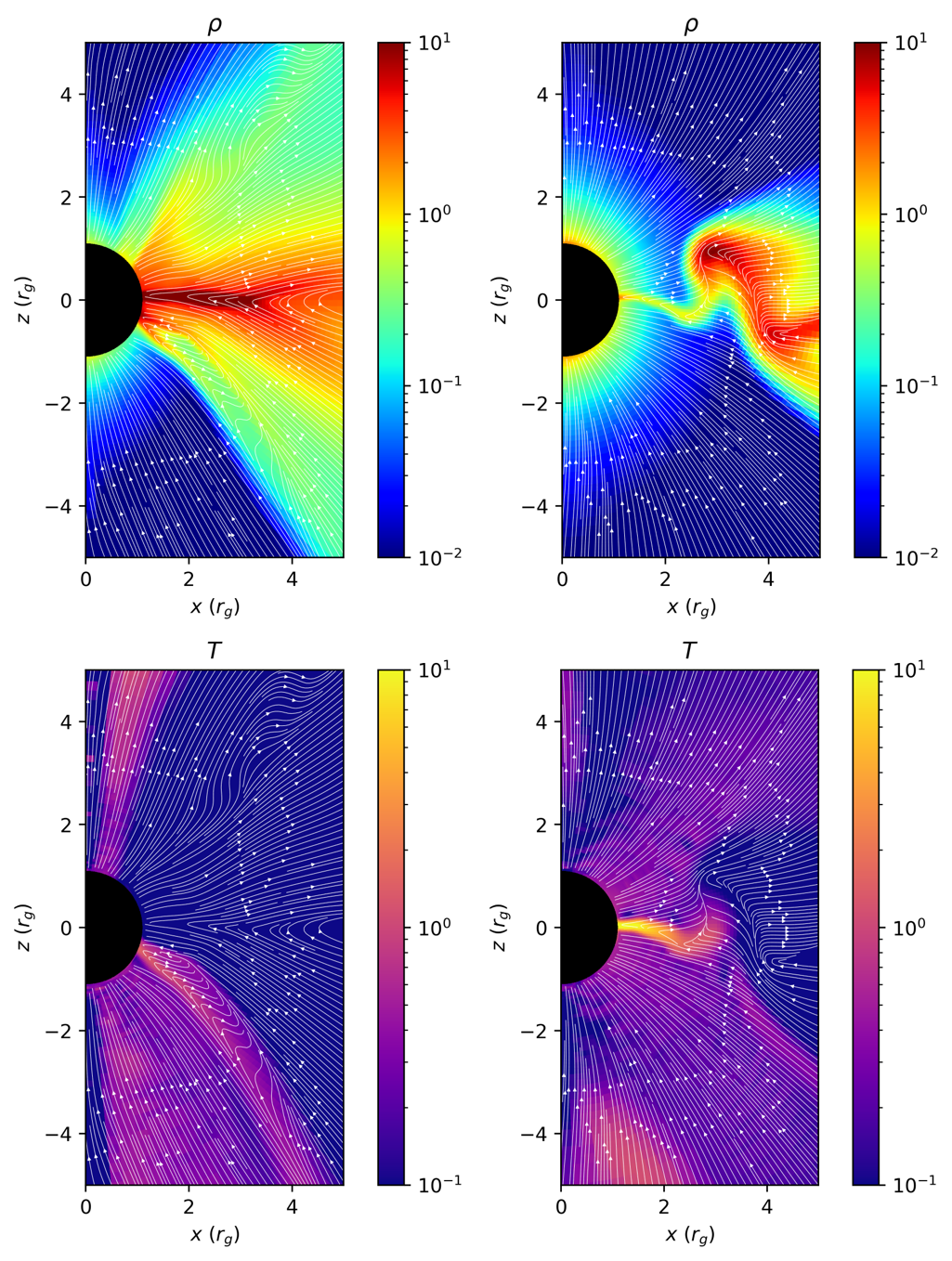}
     \caption{Distributions of plasma density (upper panels) and temperature $P/\rho$ (lower panels  at $T=25800$ (left panels) and $T=26600$ (right panels). In the left panels, $\phi_{\mathrm{BH}} = 19.5$, far below the saturation limit of MAD. Therefore the accretion flow is geometrically thick close to the balck hole. The magnetic flux of the right panels is $\phi_{\mathrm{BH}} = 49.2$, close to the saturation limit. The  accretion flow is geometrically thin due to high magnetic pressure. The temperature of the accretion flow at the equatorial plane is higher, which indicates that there should exist some heating mechanism, i.e., magnetic reconnection.}
    \label{fig:mad}
\end{figure}

Figure \ref{fig:mad} shows properties of the accretion flow from two successive moments of $t=25580$ and $t=26600$ in MAD98 simulation. At $t=25580$, the magnetic flux $\phi_{\mathrm{BH}} = 19.5$ is far below the saturation limit of the MADs. The accretion rate is relatively high ($\dot{M}=37.0$) and a geometrically thick disk structure is remarkable. The high temperature plasma are concentrated in the jet region while the accretion flow maintains relatively cool. At $t=25580$, as the magnetic flux gradually accumulates via accretion ($\phi_{\mathrm{BH}} = 49.2$) and reaches the saturation limit. In this case, the strength of the central poloidal field becomes so strong that the magnetic pressure force is comparable to the gravity, so the accretion can only occurs via exchange instability and the accretion rate drops to $\dot{M}=16.4$. At the given $\phi-$slice, the accretion flow collapses to a thin disk near the equatorial plane \citep{2012MNRAS.423.3083M}, where the gas is intensely heated by magnetic reconnection.

\subsection{Distribution of the current sheets}
\label{rcs}
In most cases, an ideal-MHD assumption is an appropriate approximation to describe the accretion flow because the plasma often have very low resistivity. The magnetic field lines can be regarded as being frozen in the fluid and moving with it. However when magnetic field lines with different polarities approach close enough, even the very weak non-ideal effects are non-negligible. The interaction of field lines will generate a current sheet according to Amp\`{e}re's law $\nabla \times \boldsymbol{B} = \mu_0 \boldsymbol{j}$ and the lines can be broken and reconnected via Ohmic dissipation within the sheet. 

\begin{figure*}	\includegraphics[width=1.\textwidth]{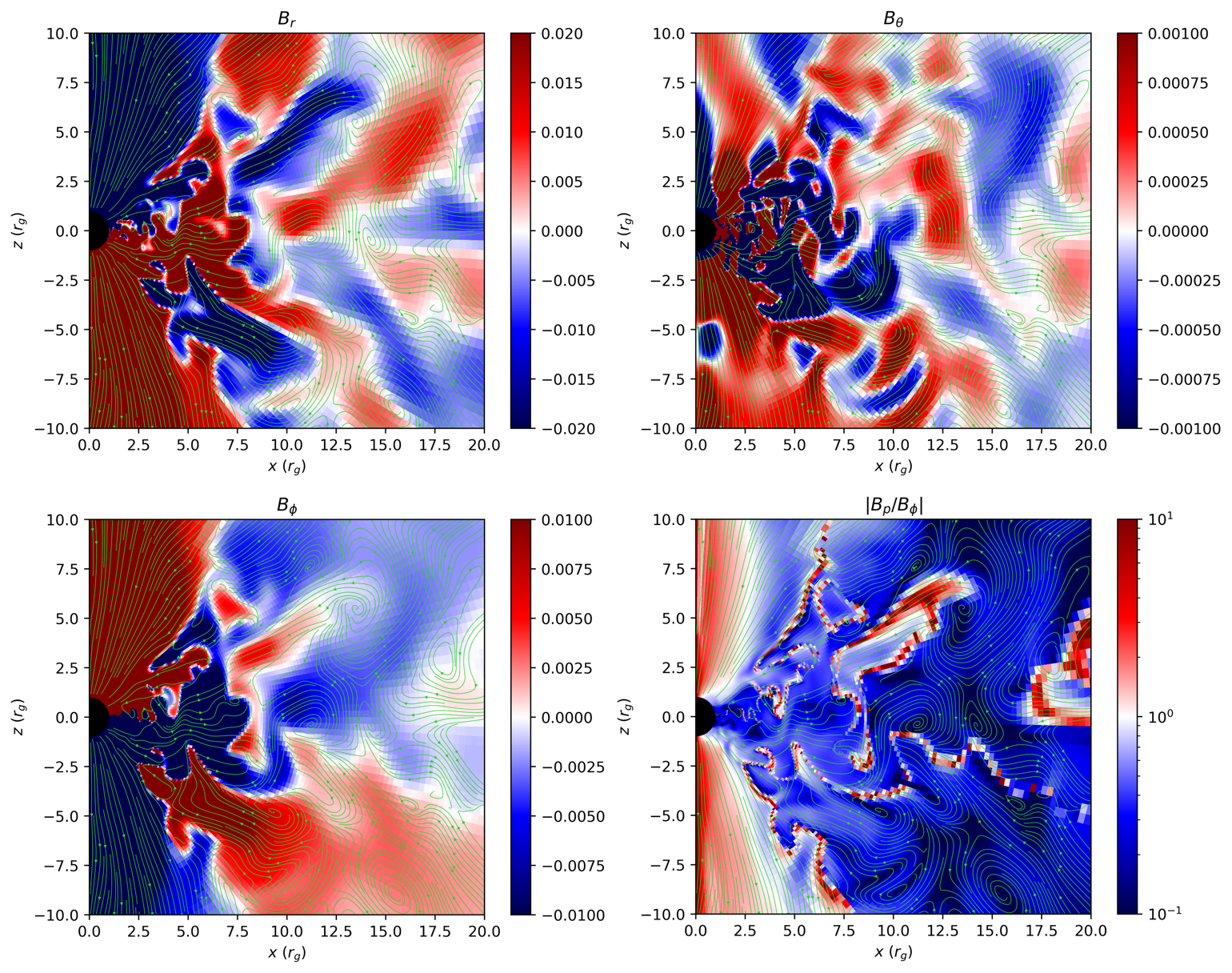}
    \caption{Distribution of various components of magnetic field in SANE98. The last panel shows distribution of the ratio of poloidal and toroidal field. It can be seen from the figure that there are many  magnetic polarity reversal regions both in the poloidal (r and $\theta$ ) and toroidal ($\phi$) directions, where  magnetic reconnection may happen.  }
    \label{fig:sane98_B}
\end{figure*}

Figure \ref{fig:sane98_B} depicts the configuration of the magnetic field for SANE98. As is shown in the first three panels, there are many regions where magnetic field reverses polarity. \citet{Ball2018b} figured out that these regions correspond to the current sheets, the potential sites of magnetic reconnection. The last panel in Figure \ref{fig:sane98_B} illustrates the ratio of poloidal and toroidal components of the field. Overall, the magnetic field of SANE98 is dominated by the toroidal field in the inner accretion flow ($|B_p/B_{\phi}|<1$) , and by the poloidal field in the jet region ($|B_p/B_{\phi}|>1$). But for some areas in the accretion flow, the poloidal field can also be overwhelming. We note that these areas coincide with most of the nulls for both $B_{r}$ and $B_{\phi}$. The takeover of $B_p$ in these areas indicates the reconnection mainly occurs on the toroidal direction there. 

\begin{figure*}	\includegraphics[width=1.\textwidth]{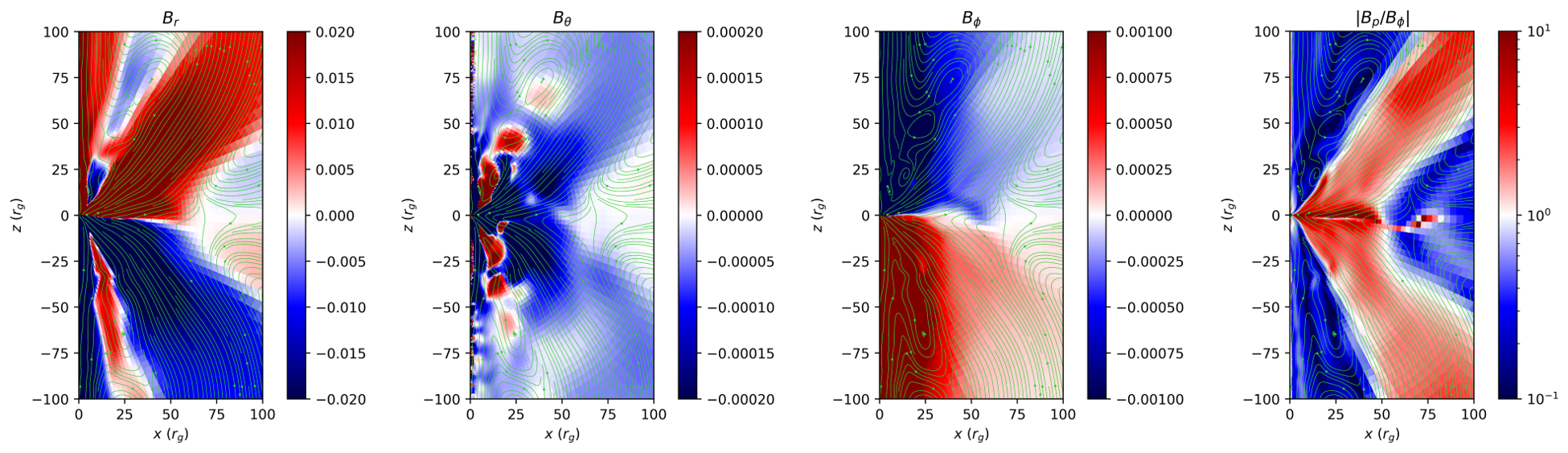}
    \caption{Same as Figure \ref{fig:sane98_B} but for MAD98. The current sheets are mainly present near the equatorial plane and jet sheath. Reconnection will therefore easily occur around the equatorial plane and result in the formation of flux ropes. Note that there exist quasi-symmetric large scale poloidal flux bundles (red region in the last panel) for the MADs. When the flux reaches a saturation value, those flux bundles can be nearly perpendicular to the equatorial plane and the corresponding magnetic force is so strong that the gas is obstructed from being accreting.}
    \label{fig:mad98_B}
\end{figure*}

Figure \ref{fig:mad98_B} shows the magnetic field configuration for MAD98. Different from SANE, current sheets are only present in the regions of the equatorial plane and jet sheath. We speculate the physical reason for the formaiton of these strong current sheets as follows. When the magnetic flux  close to the black hole becomes stronger and stronger due to the accumulation of magnetic flux, the strong toroidal magnetic field will compress the accretion flow thus the accretion flow becomes geometrically thin \citep{2012MNRAS.423.3083M}. Consequently the field lines will become aligned with the equatorial plane. The field lines with opposite polariority thus becomes very close, which produces strong current sheet. This will easily result in magnetic reconnection and the formation of flux rope, as we will discuss in the next section. The strong magnetic field in the inner accretion flow suppresses magnetorotational instability (MRI, \citealt{Balbus1991}) and subsequently turbulence driven by MRI. The small-scale current sheets are therefore more difficult to form. Since the field lines are somewhat ordered, once the reconnection occurs, it is always accompanied by the reversal of large-scale field lines. In addition, the field strength in the MADs is larger than that in the SANEs. Therefore reconnection triggered in the MADs tends to be more energetic and sometimes it could even have a great influence on the dynamics of the accretion \citep{Dexter2020}. The third panel of Figure \ref{fig:mad98_B} shows that no significant toroidal field changes sign for the jet sheath current sheets. So the reconnection at these regions is dominated by the reversal of poloidal field. As a contrast, both radial field $B_r$ and toroidal field $B_{\phi}$ switch sign for the equatorial current sheet. An apparent larger ratio of $|B_p/B_{\phi}|$ in the sheet compared to its surrounding suggest a toroidal field dominated reconnection.

\subsection{Formation and motion of flux ropes}
\label{fluxrope}

Stimulated by the distributions of the current sheet shown above, we propose the following scenario for the formation of flux ropes. 
In the case of SANE,
differential rotation and turbulent motion of the accretion flow twist the field lines, imposing different parts of the lines with opposite polarity to approach each other.
Magnetic reconnection then occurs and some plasma will be locked into the closed magnetic field lines, forming the flux rope. This is the scenario proposed in the analytical work of  \citet{Yuan2009} and confirmed later by the three dimensional GRMHD simulations by \citet{Miki2022}. 

In the case of MAD, however, the situation seems to be different. In this case, due to the suppression of MRI and turbulent motion, the reconnection seems to be mainly occurs at the equatorial plane of the accretion flow and the jet sheath, as suggested by Figure \ref{fig:mad98_B} and explained before. As the consequence of the reconnection, magnetic field lines will become closed and a flux rope will be formed. In the present paper, we focus on the flux ropes formed in the case of MAD since they are energetic enough to explain observations of the flares in Sgr A*. 

In some works, an alternative model for the formation of plasmoids has been proposed. 
In this scenario, the current sheet will become tearing-mode unstable once it becomes thin enough and its length surpasses the critical wavelength. A chain of plasmoids in various scales will then form in the sheet and fast reconnection will be simultaneously triggered \citep{Ripperda2020, Bransgrove2021}. The plasmoids are then propelled to be ejected along the sheet under magnetic tension. 
We do not find this process from our simulation data, likely because our resolution is still not high enough. We speculate that the small plasmoids formed by this process, if the process does occur, will merge with the much larger flux rope found in our simulations.  

\begin{figure*}	
\includegraphics[width=1.\textwidth]{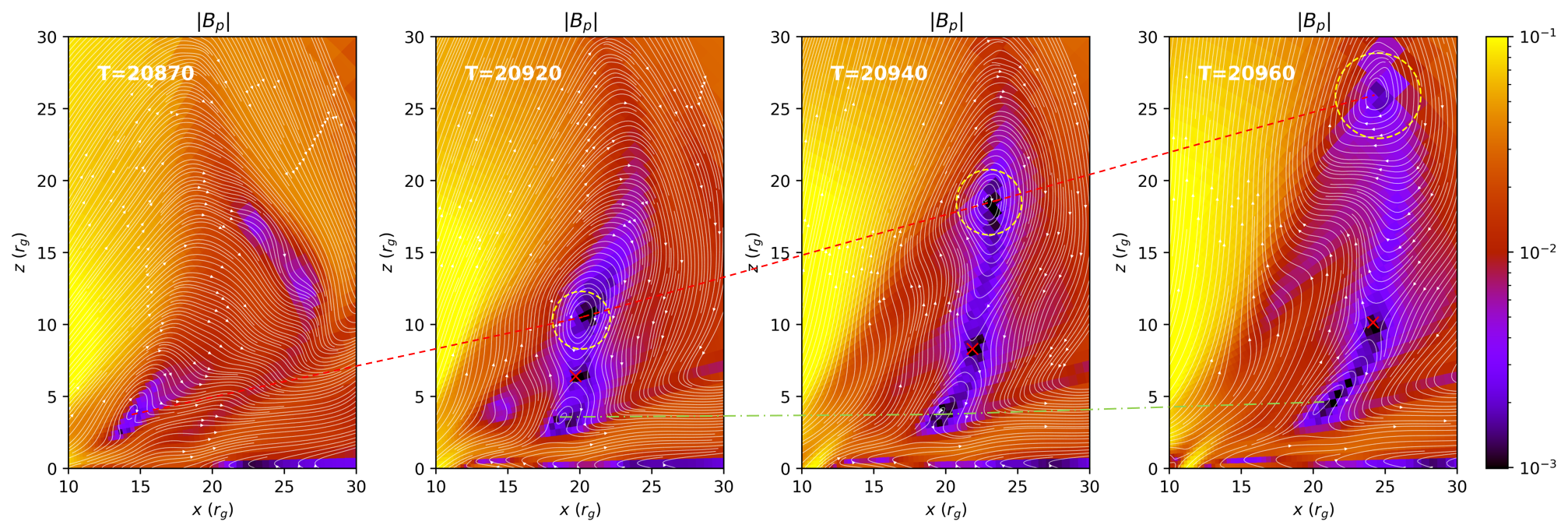}
      \caption{Evolution of a flux rope in MAD98. The leading flux rope is traced by the red dashed lines and the flare loops is traced by green dash-dotted lines. The current sheet lies between the two magnetic structures and the X point is denoted by a red cross. The yellow circle shows the expanding boundary of the flux rope blob.}
    \label{fig:rope_formation}
\end{figure*}

We have identified the flux rope from our simulation data using the approach presented in \citet{Miki2022}. In Figure \ref{fig:rope_formation}, we present the trajectory of the flux rope, indicated by red dashed lines. When the flux rope moves outward, the magnetic field lines with opposite polarities will come close enough and this is why below the flux rope there is a current sheet. The current sheet is the place where the reconnection occurs, which is also the place where electrons are accelerated. The accelerated electrons should flow into the flux rope following the reconnection outflow after they are accelerated. The radiation of these nonthermal particles is responsible for the observed flares.   Due to the continuous reconnection, the flux rope is increasing its size as it moves upward and its magnetic flux is also growing. The reconnection also helps the acceleration of the ejected flux rope, due to the enhancement of the magnetic pressure and the reducement of the magnetic tension force \citep{Yuan2009,Miki2022}. We also mark flare loops beneath the current sheet by green dash-dotted lines in Figure \ref{fig:rope_formation}. The flare loops, similar with the flux rope, are also the congregation of the magnetic field lines after being reconnected. They suffers from intense shearing and grows slowly compared to the flux rope \citep{Gou2019}. A small portion of the electrons accelerated in the current sheet will also flow into this area, which are nevertheless neglected by us in this work since their radiation is likely weaker \citep{Lin2023}. 

In the case of SANE, the flux ropes are distributed widely in the accretion flow, similar with the current sheets. There are many flux ropes presenting at the same time. The turbulence in the flow makes the motion of the flux rope more turbulent. In the case of MAD, most of the flux ropes are initially formed around the equatorial plane and then ejected into the corona region. The flux rope propagate outward along the equator if it is formed in the equatorial plane. Because the accretion flow in MAD98 is less turbulent, the movement of the flux rope in the radial direction is straighter compared to the case of SANE. Generally, only one or two flux ropes are found to exist at a time in MAD98. 

The above-mentioned different distribution of flux ropes in the cases of SANE and MAD will be reflected on radiation if we assume the non-thermal electrons are concentrated on the flux rope center. In the case of MAD, the radiative plasma blob will appear in the form of ``hot spot'', which can well explain the centroid motion and polarization angle swing if it rotates to the black hole \citep{Broderick2006,GRAVITY2018,GRAVITY2020orb}. However, due to the widely stochastic distribution of the current sheets and flux ropes in SANE98, it can be imaged that the flux centroid of these hot spots will be very close to the central black hole and it is very hard for them to reproduce an apparent rotation pattern unless there is an emission dominating flux rope. Because of these reasons, we focus on the flux rope generated in the MAD98 and investigate whether we can explain GRAVITY observations based on our ejective flux rope model. 

\begin{figure}
	\includegraphics[width=\columnwidth]{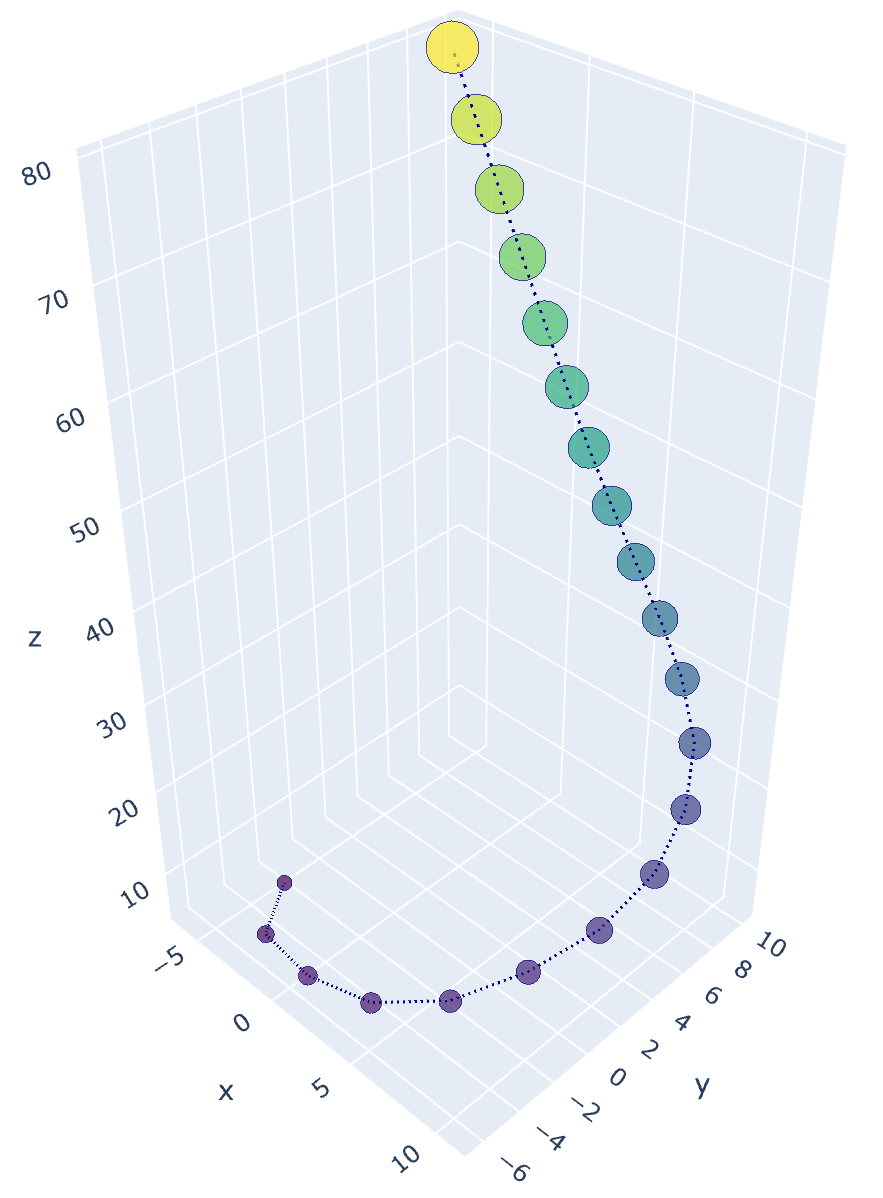}
    \caption{Trajectory of the center of a flux rope lasting for $t\sim 200r_g/c$ in MAD98 (color ranging from dark purple to yellow). The increasing size of the dots indicate the flux rope is expanding through time.}
    \label{fig:trajectory}
\end{figure}

\begin{figure}
	\includegraphics[width=\columnwidth]{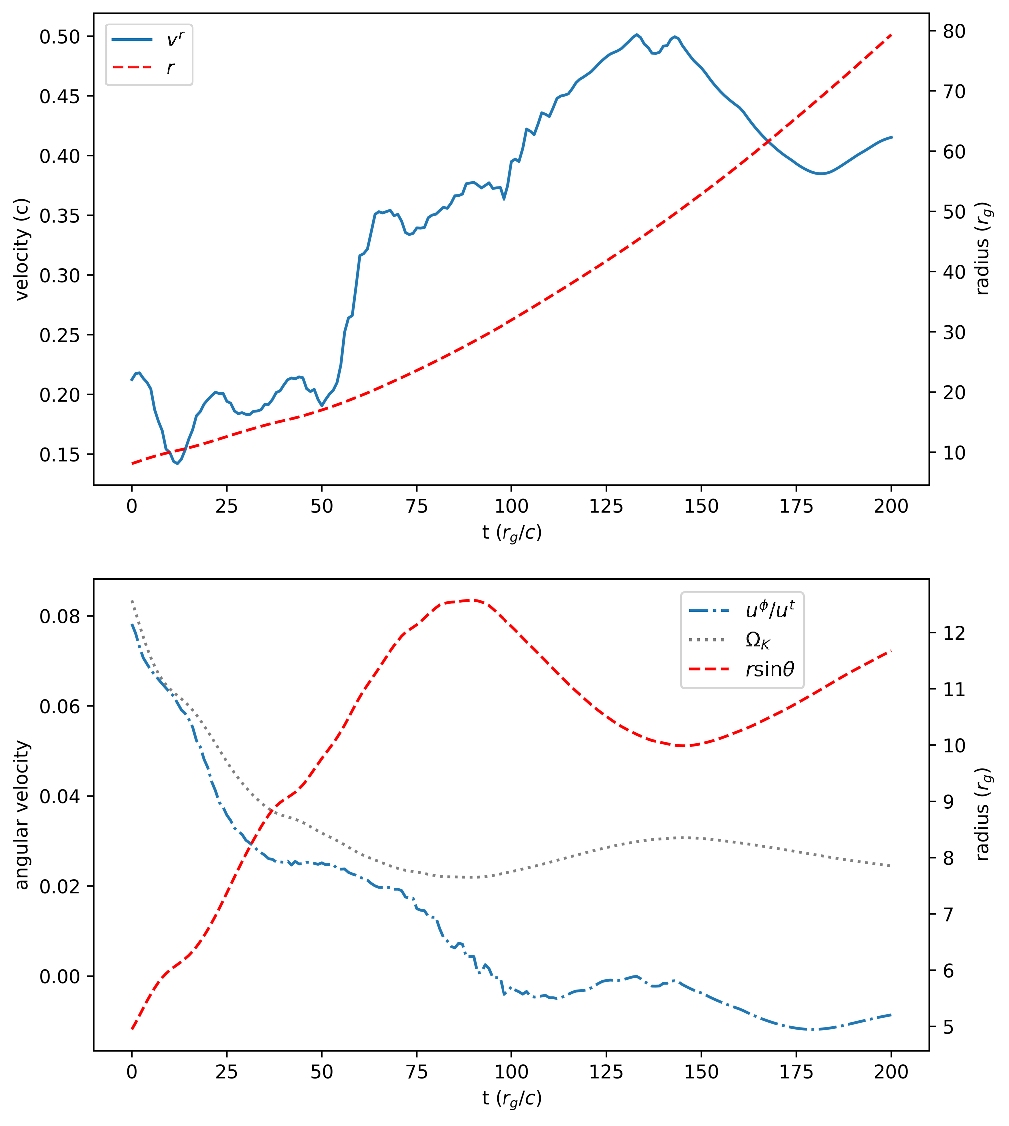}
    \caption{The velocity of the flux rope. Upper panel shows the radial velocity and distance to the black hole. Lower panel show the angular velocity in comparison with the local Kepelarian velocity as well as the distance to the spin axis.}
    \label{fig:velocity}
\end{figure}

In the 3D simulations, the flux rope is extended in the azimuthal direction \citep{Miki2022}. To make the problem simple, in our calculations we hypothesize that the velocity of the flux rope is identified with the fluid velocity at the center of the rope and regard the flux rope moves with an identical velocity.  We can then reconstruct rope's trajectory by integrating the velocity over time according to our simulation data once we know its position at a moment by
\begin{equation}
\frac{\mathrm{d} x_{c}^\mu}{\mathrm{d} t}=\frac{u_{c}^\mu}{u_{c}^0},
\end{equation}
where $x_{c}^{\mu}$ and $u_{c}^{\mu}$ are the coordinates and fluid four-velocity at the flux rope center \citep{2015ApJ...804..101Y}. A fourth-order Runge–Kutta method is adopted to do the integration. Figure \ref{fig:trajectory} shows the motion of a flux rope initially formed at $t=25800r_g/c$, $r = 8.1 r_g$, $\theta=37.6^{\circ}$ and $\phi=213.8^{\circ}$ in MAD98. The flux rope is moving outward while rotating around the spin axis, exhibiting a helical motion. It is accelerated very efficiently and quickly reaches a relativistic velocity of $\sim 0.5 c$ and a height of $80r_g$ within $200r_g/c$, as shown by Figure \ref{fig:velocity}. We also compare the angular velocity of the fluid in the flux rope center with the local Keplerian velocity.  Those two velocities are very close when the flux rope is not too far away from the black hole ($r \lesssim 25r_g$). 

When tracing the motion the flux rope, we find sometimes it will be accreted into the black hole rather than ejected outward. \citet{Bransgrove2021} indicated that generally plasmoids formed in the equatorial current sheets will move towards the black hole with slow velocities ($v<c$) if they are born inside the stagnation surface, but will be ejected out and accelerate to relativistic velocities if born outside the surface. \citet{Miki2022} studied flux ropes generated in the corona region and obtained a similar result. The separatrix in their simulation is located at $r \sim 10-15r_{\mathrm{g}}$. Our more careful analysis indicates that  the saperatix not only depends on the distance from the black hole, but also on the longitude. Generally, the rope generated near the polar region is easier to be ejected out with a higher velocity than which is formed near the mid-plane. 

Because the trajectory method is computational consuming and hard to be adopted to the overall simulation region, we instead test whether a fluid element can be ejected out by employing geodesic criterion, i.e., if the time component of its covariant four-velocity satisfies $u_t \leq -1$ \citep{Bovard2017}. This criterion is equivalent for the fluid element to have non-zero velocity at infinity, if they strictly follow the geodesic motion.
It should be emphasized that this criterion has limitation because the fluid elements experience other forces, i.e., the magnetic forces and gas pressure, when they move outward so that they do not strictly follow the geodesics. This situation is similar to another criterion adopted in the literature, i.e., the Bernoulli parameter criterion. In some works, authors judge whether the fluid elements can escape to infinity by evaluating the value of Bernoulli parameter $hu_t \leq -1$\citep[e.g.,][]{Bovard2017}. However, as shown by \citet{2015ApJ...804..101Y} using the trajectory approach, since the accretion flow is not strictly steady and inviscid, Bernoulli parameter is not a constant of motion. Nevertheless this criterion provides a simple picture of where the flux ropes are more easily to be ejected out to infinity. 

\begin{figure*}	
\includegraphics[width=1.\textwidth]{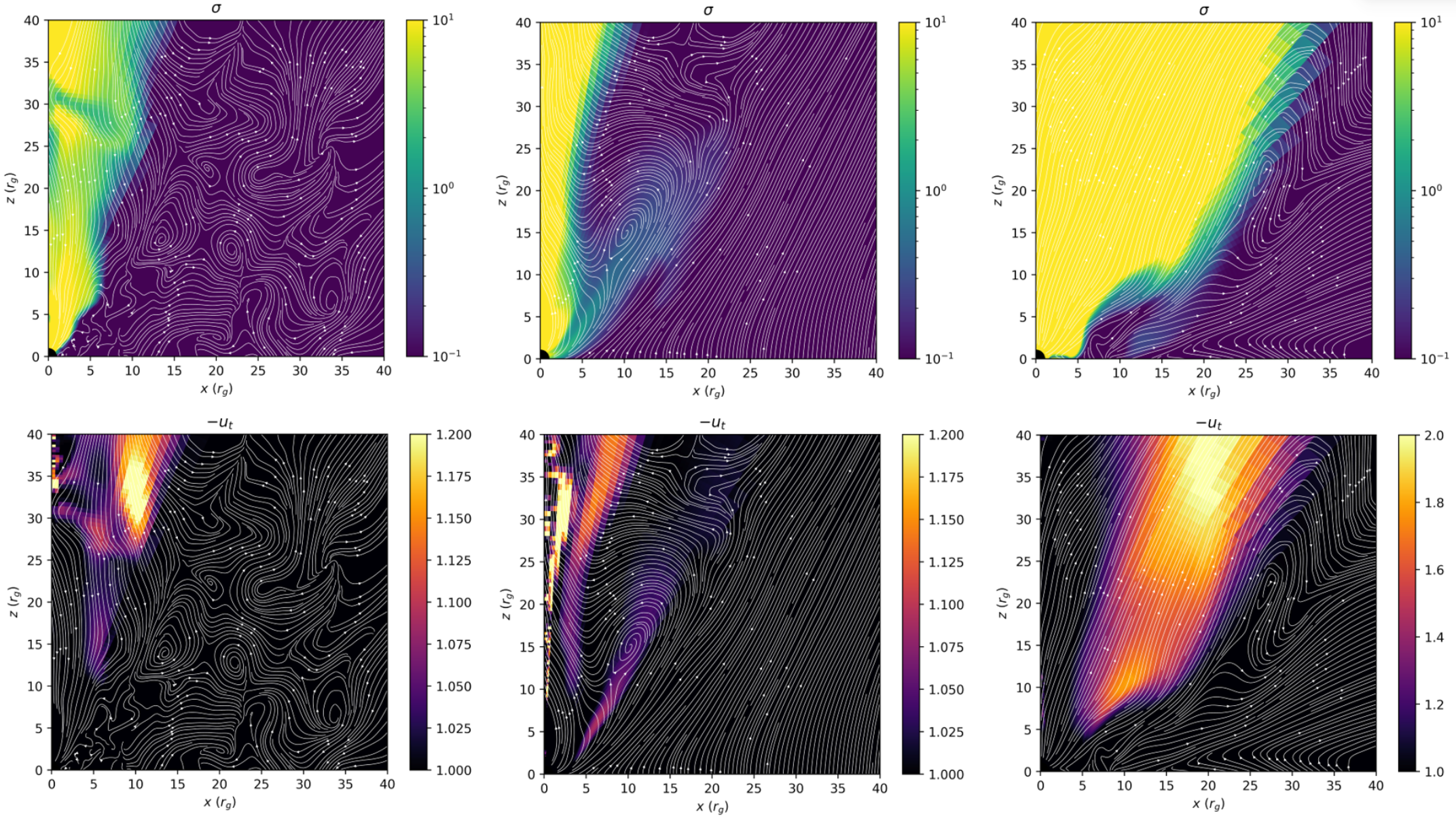}
    \caption{Plasma $\sigma$ and $-u_t$ in SANE98 (left panel), accretion state (middle panel) and halt state (right panel) in MAD98. }
    \label{fig:sigma_ut}
\end{figure*}

Figure \ref{fig:sigma_ut} shows the plasma $\sigma=b^2/\rho$ and $u_t$ of SANE98 ($T=26600$; the left panel), and the low ($T=25850$; middle panel) and high ($T=26600$; the right panel) magnetic flux states of MAD98. The mass accretion rates are high and low in the latter two cases. From the lower panels of this figure, we see that close to the jet region, the flux ropes are more likely to be ejected out. The flux ropes near the equatorial plane, however, seem to be difficult to escape from the black hole's gravity.  In all three cases, the unbound regions are located in the high-magnetized regions. This indicates that magnetic forces play an important role in the process of fluid to resist gravity. 
The high-magnetized and unbound plasma are concentrated on the jet-wind boundary region. As the acceleration of the electrons are more efficient in the high $\sigma$, low $\beta$ region \citep{Ball2018b,Nathanail2020}, the current sheets and flux ropes near the jet are preferred to make major contribution to the flux during flare events.  Thus the flux ropes, which are prone to be formed between the jet and accretion flow regions, are more close to the spin axis.


\section{Determination of nonthermal electrons}
\label{sec:nonthermalelectrons}

\subsection{Production of non-thermal electrons by reconnection}
The reconnection of the field and acceleration of the particles can 
only be accurately studied when kinetic particle-in-cell (PIC) methods are employed \citep{Ball2018b,Mellah2023}. In this work, we follow \citet{Li2017} and \citet{Lin2023} to use a semi-analytical phenomenological  model to determine the injection and evolution of the non-thermal electrons in the flux rope that are supposed to make the most contribution to the flare flux \citep{GRAVITY2018}. Combined with the magnetic field and velocity field obtained from the simulations, we can then conduct relativistic ray tracing radiative transfer calculations. We first convert simulation units to cgs units by hypothesizing that the average accretion rate  is $\dot{M} = 5 \times 10^{-8} {M_\mathrm{\odot}}\mathrm{yr}^{-1}$. This value satisfies the constraint that $2 \times 10^{-9} \mathrm{M}_{\odot} \mathrm{yr}^{-1}<\dot{M}<2 \times 10^{-7} \mathrm{M}_{\odot} \mathrm{yr}^{-1}$ based on rotation measure \citep{Marrone2007,Wang2013} . 

Magnetic reconnection is an effective approach for the electrons initially in thermal distribution to gain energy and transit from Maxwell-J{\"u}ttner distribution to power-law distribution.
\citet{Ball2018a} investigated electron and proton acceleration in the trans-relativistic regime by PIC simulations. They deduced the dependence of the power-law slope and acceleration
efficiency on the ambient plasma-$\beta$ and magnetization $\sigma$, which was adopted to explain Sgr A* flares in later works \citep{Chatterjee2021,Aimar2023}. Stimulated by the PIC simulation results, \citet{Petersen2020} suggested that the non-thermal electron acceleration rate is proportional to the $J^2$, the square of the current density. This prescription was adopted most recently by \citet{Yang2024} to investigate the acceleration and energy distribution of nonthermal electrons in jets. 

In the present work, we assume electrons are energized by the energy flux released in the reconnection process. The total injection power is given by
\begin{equation}
P_{\mathrm{inj}}=\iint \mathbf{G} \cdot \mathrm{d} \mathbf{A},
\label{eq:injection power}
\end{equation}
where $\mathbf{G}$ is the energy flux and the integral region is the area of the reconnecting current layer. In the simplest ideal MHD approximation where the current layer is magnetic neutral, the energy flux density can be expressed as
\begin{equation}
\label{eq:energy flux}
\mathbf{G}=\rho \mathbf{v}\left(\frac{v^2}{2}+w\right)+\frac{1}{4 \pi}\left(B^2 \mathbf{v}-(\mathbf{B} \cdot \mathbf{v}) \mathbf{B}\right),
\end{equation}
where $w=\epsilon+p/\rho$ is the specific enthalpy. We further assume that the magnetic field plays a dominant role in the energy supply, which is reasonable since a strong-field approximation is satisfied in the place where reconnection occurs. Then the second term, i.e., $\mathbf{G}_{\mathrm{p}}=\frac{1}{4 \pi}\left(B^2 \mathbf{v}-(\mathbf{B} \cdot \mathbf{v}) \mathbf{B}\right)$, which is known as magnetic energy flux density or the Poynting vector, is dominating. Near the current layer, the magnetic field can be approximately regarded as perpendicular to the plasma velocity $\mathbf{B} \perp \mathbf{v}$. Therefore the Poynting vector will be treated as flowing directed into the the current layer, i.e.,

\begin{equation}
\mathbf{G}_{\mathrm{p}}=\frac{1}{4 \pi} B^2 \mathbf{v},
\end{equation}
and the total inflow energy injection rate is 
\begin{equation}
P_{\mathrm{inj}}=\iint \frac{1}{4 \pi}B_{0}^2 {v_0} \ \mathrm{d} {A},
\end{equation}
where $B_0$ is the field strength in the inflow region and $v_0$ is the inflow velocity. Since the MAD current sheets are typically flat and regular in shape as shown in Figure \ref{fig:mad98_B}, we simply assume these current sheets are cone-shaped surfaces extending in the poloidal and azimuthal directions \citep{Mellah2023}. The center of the current sheet can be identified with the X point. And the length $L$ of it in the poloidal direction is assumed to be proportional (1/10) to the distance between flux rope center and the field foot point, which is inspired from our previous analytic work \citep{Li2017,Lin2023}. The angle that the layer stretches in the azimuthal direction is fixed to be $\phi_{\mathrm{cs}}=100^{\circ}$, as suggested by the numerical simulations of \citet{Miki2022}. The inflow region is chosen to be $1r_g$ \footnote{It is adopted to be $0.5 r_g$ in \citep{Ball2018b}.} away from where the poloidal magnetic field reaches a local minimum (i.e., the current layer center). At this distance the poloidal magnetic field also roughly reaches an asymptotic value \citep{Ball2018b}. We sample the magnetic field here and assume a constant injection rate $v_0$ = 0.1 $V_A$, where $V_A = \sqrt{(\sigma/(\sigma+1))}\ c$ is the Alfv{\'e}n velocity which can be directly calculated based on the simulation data.





Our target is to give a reasonable injection power according to the simulation data. Since it is far from trivial to fully identify and characterize current layers in the GRMHD simulation, not to say integrating over the surface, we have to make some assumptions and simplifications. The relativistic effects are neglected. The injected magnetic energy will then be converted into the electrons' energy via Joule dissipation in the current sheet. A part of the electrons will be  accelerated into non-thermal power-law distribution. Then the injection rate for the electrons with energy $\gamma$ is
\begin{equation}
n_{\mathrm{pl}}(\gamma)=c_{\mathrm{inj}}^{\prime} \gamma^{-p_{\mathrm{e}}}, \quad \gamma_{\min } \leq \gamma \leq \gamma_{\max },
\label{eq:power law}
\end{equation}
where $\gamma_{\min}$ and $\gamma_{\max}$ are the minimum and maximum electron Lorentz factors, respectively, $p_{\mathrm{e}}$ is the power law index and $c_{\mathrm{inj}}^{\prime}$ is the normalization. $\gamma_{\min}$ can be determined if we assume it is tied to the peak of the Maxwellian distribution of the thermal electrons:
\begin{equation}
\gamma_{\min }=\gamma_{\mathrm{pk}} \simeq 1+u_{\mathrm{th}} / m_{\mathrm{e}} c^2,
\end{equation}
where $u_{\mathrm{th}}=f\left(\Theta_{\mathrm{e}}\right) \Theta_{\mathrm{e}} m_{\mathrm{e}} c^2$ is the energy density of the thermal electrons. Here $\Theta_{\mathrm{e}} = k_{\mathrm{B}} T_{\mathrm{e}} / m_{\mathrm{e}} c^2$ is the dimensionless thermal temperature and $f\left(\Theta_{\mathrm{e}}\right)=\frac{6+15 \Theta_{\mathrm{e}}}{4+5 \Theta_{\mathrm{e}}}$ \citep{Gammie1998}. GRMHD simulations provide us the proton temperature $T_\mathrm{p}$. Following \citet{Monika2016}, the electron temperature $T_{\mathrm{e}}$ can be related to $T_{\mathrm{p}}$ and plasma-$\beta$ by
\begin{equation}
\frac{T_{\mathrm{p}}}{T_{\mathrm{e}}}=R_{\mathrm{high}} \frac{\beta^2}{1+\beta^2}+R_{\mathrm{low}} \frac{1}{1+\beta^2}
\end{equation}
where $\beta=P_\mathrm{g}/P_\mathrm{B}$, $R_{\text {low}}$ and $R_{\text {high}}$ are the parameters that measure $T_{\mathrm{p}} / T_{\mathrm{e}}$ in the low and high plasma-$\beta$ regions. We choose $R_{\text {low}}=1$ and $R_{\text {high}}=100$. The exact value of maximum Lorentz factor $\gamma_{\max}$ is not so important as long as it is large enough ($\gtrsim 10^5$) to power X-ray flare \citep{Markoff2001}. In this work we fix $\gamma_{\max} = 10^6$. The magnetic reconnection provides injection power according to Equation \ref{eq:injection power}. We assume a part of this energy is transformed to non-thermal electrons, namely $u_{\mathrm{nth}}^{\prime} = \eta P_{\mathrm{inj}}$, where $0\leq \eta \leq 1$ is the non-thermal acceleration efficiency. Hence

\begin{equation}
\eta P_{\mathrm{inj}}=\int_{\gamma_{\min }}^{\gamma_{\max }} \gamma c_{\mathrm{inj}}^{\prime} \gamma^{-p_{\mathrm{e}}} m_{\mathrm{e}} c^2 \mathrm{~d} \gamma.
\label{eq:efficiency}
\end{equation}
If we assume a constant injected power-law index $p_{\mathrm{e}}$, then we can deduce $c_{\mathrm{inj}}^{\prime}$ by solving Equation \ref{eq:efficiency}. The specific value of the model parameters are summarized in Table \ref{tab:electron parameters}.
\begin{table}
\centering
\caption{Parameters for the non-thermal electrons.}
\begin{tabular}{lccccc}
\hline \hline Parameter & Value \\
\hline
injection power-law index $p_{\mathrm{e}}$ & 2.3 \\
maximum Lorentz factor $\gamma_{\mathrm{max}}$ & $10^6$ \\
acceleration efficiency $\eta$ & $10\%$ \\
initial blob radius $R_0 (r_g)$ & 2.0 \\
blob expansion velocity $v_{\mathrm{exp}} (R_0/\mathrm{hr})$ & 1.0 \\
\hline
\label{tab:electron parameters}
\end{tabular}
\end{table}
We note that the electron injection term is not calculated exactly on every grid. The MAD current sheet near the jet sheath is quite straight and distribute along the radial direction at a constant $\phi$-slice. The physical quantities tend to vary smoothly and uniformly along the sheet as well. Therefore we average the quantities along the current sheet where the sheet structure is most apparent, and assume the obtained values to be approximately suitable for other azimuthal angles. Then the integration can be reduced to
\begin{equation}
P_{\mathrm{inj}} \approx \frac{1}{4 \pi} \langle B_{0} \rangle^2 \langle {v_0} \rangle  L R_c \phi_{\mathrm{cs}},
\end{equation}
where $\langle B_{0} \rangle$ and $\langle {v_0} \rangle$ are the averaged magnetic field in the reconnection region and averaged inflow velocity, respectively, and $R_c$ is the radial coordinate of the current sheet center. 

\subsection{Evolution of the nonthermal electrons}

Non-thermal electrons accelerated in the current sheet should not stay in the acceleration site. Rather they should flow out along the current sheet together with the reconnection outflow and converge at the  flux rope. The outflow velocity is Alfv{\'e}n speed $\sim V_A \sim c$ in the high $\sigma$ region, with $V_A$ being the Alfv{\'e}n speed. The maximum length scale of the current sheet is of $\sim 10 r_g$. Therefore the accelerated electrons will flow into the flux rope region within $10r_g/c$ or $\sim3.5$ mins when scaled to the mass of Sgr A*. On the other hand, the synchrotron cooling timescale for the electrons is $1.29 \times 10^{12}/(\nu^{1/2} B^{3/2})$ s. The typical magnetic field strength in the current sheet is roughly an order lower than the ambient flow, so the field strength in the current sheet after reconnection is $\lesssim$10 G. Given that the frequency of the NIR radiation is $\sim 1.36
\times 10^{14}$ Hz, the typical cooling timescale in the current sheet is estimated to be longer than an hour. In other words, these accelerated electrons will rapidly flow into the flux rope region before they consume most of their energy by synchrotron radiation. Therefore we neglect electrons' travelling time in the current sheet and assume most of the radiation is directly emitted from the flux rope region. 

The flux rope, as the concentration of the reconnected magnetic field lines, has a large field strength. The radiation cooling effect in this region is thus non-negligible. We also take into account the effect of adiabatic expansion of the blob, assuming the accelerated electrons are gathered in a spherical area in the center of the flux rope. The expansion effect is crucial for explaining the time-delay between the peaks of muti-wavelength radio flares \citep{Zadeh2006,Zadeh2009}. For instance, VLA observation has found that the peak flare emission at 43 GHz leading the 22 GHz flare by $\sim20 - 40$ minutes. This has been explained as arising from the emitting blob changing from optically thick to optically thin due to expansion in the process of ascending \citep{Zadeh2006}. 

By considering the radiative and adiabatic cooling processes, the time-dependent non-thermal electron distribution function $N(\gamma,t)$ can finally be determined according to the continuity equation \citep{Dodds-Eden2010,Li2017}:
\begin{equation}
    \frac{\partial N_{\mathrm{e}}(\gamma, t)}{\partial t}=Q_{\mathrm{inj}}(\gamma, t)-\frac{\partial\left[\dot{\gamma} N_{\mathrm{e}}(\gamma, t)\right]}{\partial \gamma},
    \label{nonthermale}
\end{equation}
where $N_{\mathrm{e}}(\gamma,t)$ is the electron distribution, $Q_\mathrm{inj}(\gamma,t)$ is the injection term which can be written in the power-law form $Q_{\mathrm{inj}}(\gamma,t)=n_{\mathrm{pl}}(\gamma,t)=c^{\prime}_{\mathrm{inj}}(t)\gamma^{-p_{\mathrm{e}}}$ as in Equation \ref{eq:power law}. And $\dot{\gamma}=\dot{\gamma}_{\text {syn }}+\dot{\gamma}_{\text {ad }}$ are the synchrtron radiation and adiabatic expansion cooling rate, respectively. The sychrotron cooling term is
\begin{equation}
\dot{\gamma}_{\mathrm{syn}}=-\gamma / t_{\mathrm{syn}}
\end{equation}
where $t_{\text {syn }}=7.7462 \times 10^8 /\left(\gamma B^2\right) \mathrm{s}$ is the synchrotron cooling timescale. The expansion cooling term is
\begin{equation}
\dot{\gamma}_{\mathrm{ad}}=-\gamma d \log R / d t=-\gamma v_{\exp } / R
\end{equation}
where $R$ is the size of the expanding flux rope blob and $v_{\exp }=dR/dt$ is the expansion velocity. Here we assume all the non-thermal electrons are normally distributed in an expanding spherical Gaussian-like blob to mimic the plasmoid for simplification, while in the simulation it could have extension in the azimuthal direction. The size of the blob is hard to determine explicitly. Therefore we adopt a constant expansion rate $v_{\mathrm{exp}} = 1R_0/\mathrm{hr}$, where $R_0$ is the initial size of the flux rope blob, which is set to be $R_0=2r_{\mathrm{g}}$ \citep{Dodds-Eden2010,GRAVITY2020pol}. The circles in Figure \ref{fig:rope_formation} show the edges of the flux ropes obtained from tracing the trajectories of the flux rope fluid elements. The expansion of the flux rope is evident. We have compared this expansion result with that calculated using the assumed constant expansion rate and found that they differ only by a factor of $\sim 2$.  The temporal non-thermal electron distribution can be integrated explicitly by using Green function \citep{Dermer2009,Syrovatskii1959}.

\section{Interpreting GRAVITY observations of Sgr A* flares}
\label{sec:results}
After we have obtained the evolution of magnetic field and nonthermal electrons, we use a general relativistic ray-tracing radiative transfer code {\tt GRTRANS} \citep{Dexter2009,Dexter2016} to calculate all four Stokes parameters and position of the flux centroid on the sky plane. We adopt the "fast light" treatment in which the light traveling time is neglected to save computation time. We assume that the emission from Sgr A* flare is essentially contributed mainly by  non-thermal electrons in the flux rope, although the absorption and Faraday rotation caused by thermal electrons have been taken into account. The contribution from thermal electrons is neglected, partly because the temperature of the plasma in the highly magnetized region of the jet is hard to be determined precisely in GRMHD simulations. Since the NIR flux during flares is typically $\sim 10$ times larger than quiescent state, we can also safely assume that the centroid and polarization angle traces the motion of the emitting flux rope. The values of model parameters for the ray-tracing calculation are listed in Table \ref{tab:ray-tracing}. The final images have been clockwise rotated $120^\circ$ to match the observation. 

To summarize, the quantities which are taken directly from the GRMHD simulation data include the gas density, gas pressure, gas velocity field and magnetic field. The position of the flux rope blob can be determined by the velocity field. The free parameters in Table \ref{tab:electron parameters} are introduced to determine the distribution of non-thermal electrons contained within the flux rope blob. Once we have information of the non-thermal electrons as well as the magnetic field, we can make ray-tracing and radiative transfer calculations with the help of the additional parameters listed in Table \ref{tab:ray-tracing}. 
\begin{table}
\centering
\caption{Parameters for the ray-tracing calculation.}
\begin{tabular}{lccccc}
\hline \hline Parameter & Value \\
\hline
mass $M_{\mathrm{BH}} (M_\mathrm{\odot})$  & 4.1 $\times 10^6$ \\
distance $D \mathrm{(kpc)}$ & 8.1 \\
spin $a$ & 0.98 \\
inclination angle $i (^\circ)$ & 173 \\
screen resolution & $200 \times 200$\\
screen size ($\mathrm{\upmu as} \times \mathrm{\upmu as}$) & $[-200,200] \times [-200,200] $ \\
\hline
\label{tab:ray-tracing}
\end{tabular}
\end{table}

\subsection{Kinetic motion of the flux rope and comparison of its trajectory with the GRAVITY results}

\begin{figure}
	\includegraphics[width=\columnwidth]{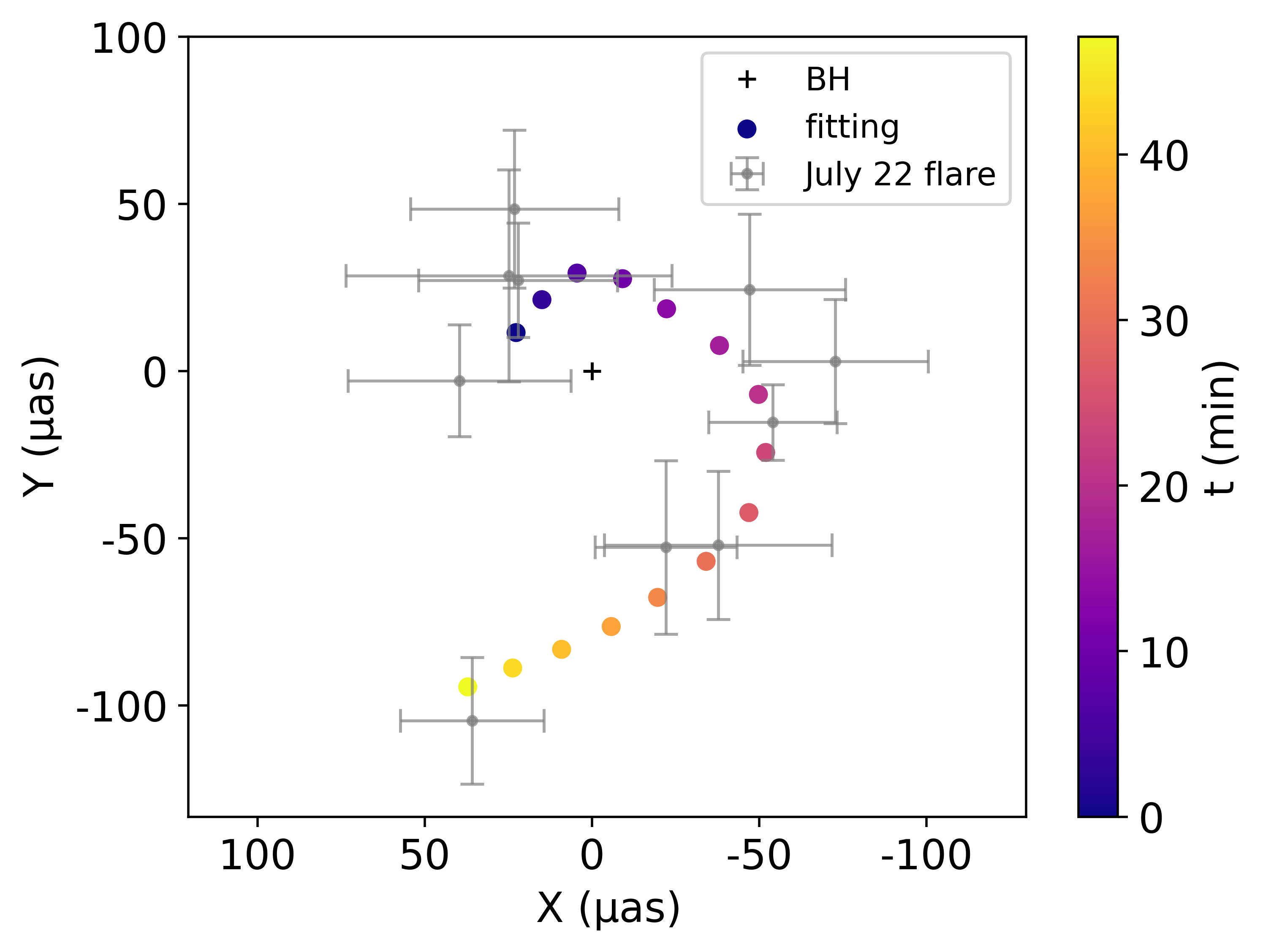}
    \caption{The trajectory of the brightness centroid in 47min. The centroid rotates continually with increasing radii. The data points with error bar is from July 22 flare.}
    \label{fig:centroid}
\end{figure}

In Figure \ref{fig:centroid}, we show the trajectory of the brightness centroid projected on the sky plane based on ray-tracing calculations and its comparison with the observation of the July 22 flare. The motion presented in the figure lasts for $140r_{\mathrm{g}}/c \approx 47$ min, which is also the case for the following calculation results of the lightcurve and polarization. With the small inclination angle, we have successfully reproduced a circular trajectory with increasing radii with time, which reflects the helical motion of the flux rope in space (refer to Figure \ref{fig:trajectory}). In \citet{GRAVITY2018}, the astrometric of all three flares exhibits ``hot spot'' of more than 100 $\mathrm{\upmu}$as $(\approx 20r_{\mathrm{g}})$ away from the central black hole. If the hot spot is located on the equatorial plane and rotates with Keplerian velocity, such a distance corresponds to a rotation period of $\sim 190$ min, much longer than the observed duration and orbital period of the flares which is typically around 60min. In our model, the distance of the ejected flux rope to the black hole spin axis (i.e., the cylindrical radius of the flux rope) ranges from $5r_{\mathrm{g}}$ to $12.5 r_{\mathrm{g}}$. This corresponds to an orbital period of $\sim60$ min, which is roughly consistent with observation. The much larger projected distance of the hot spot to the black hole shown in Figure \ref{fig:centroid} is due to the contribution of the height of the flux rope to the equatorial plane, which can be as large as 60$r_{\mathrm{g}}$. 

It is also needed to note that the July 22 flare exhibit super-Keplerian motion on the projected plane. This is explained in \citet{Lin2023} by assuming that the field lines connecting the flux rope to the underlying accretion flow is rigidly rotating. In our simulation, however, we did not find fluid velocity apparently exceed the local Kepler velocity (lower panel of Figure \ref{fig:velocity}), but slightly lower than Keplerian value when $r$ is not too large. The reason for the super-Keplerian motion is related to the above-mentioned projection effect. 
Because the projected distance of the hot spot is much larger than its cylindrical radius, the ``apparent'' angular velocity of the hot spot in unit of the Keplerian velocity at the projected distance in the projection plane becomes much larger than its ``real'' angular velocity in unit of the Keplerian velocity at its cylindrical radius. In fact, the angular velocity in unit of the Keplerian velocity at the radius of projected distance is $\sim0.96$.


This value is still slightly smaller than the observed result of super-Keplerian rotation of the hot spot. Another contribution to the super-Keplerian motion is ``light aberration'' effect.
Throughout our ray-tracing calculations, we have neglected light travelling time. This simplification can significantly affect the observed angular velocity of the ejected flux ropes. 
Considering a hot spot rotating around the black hole's spin axis with a constant radius $R$ and constant angular velocity $\omega$, while also moving radially with a velocity $v=\beta c$ at an angle $\theta$ from the observer's line of sight. For simplicity let's neglect GR effects and special relativistic effects caused by the rotation. The maximum observed rotation angular velocity can be approximated by $\frac{v \mathrm{sin} \theta+\omega R \mathrm{cos}\theta }{\Gamma(1-\beta \mathrm{cos} \theta)}/ R\mathrm{cos} \theta$, here $\Gamma=1/(1-\beta^2)$ is the bulk Lorentz factor. 
Adopting typical values of $\theta \sim 10^{\circ}$, $v \sim 0.25c$, $\omega \sim 0.02 c/r_g$ and $R \sim 10  r_g$ adopted from our simulation, the observed angular velocity of the blob will be $\sim 0.031 c/r_g$, which is $\sim 1.55$ times of the ``original'' angular velocity.

\subsection{Light curves}
\begin{figure}
	\includegraphics[width=\columnwidth]{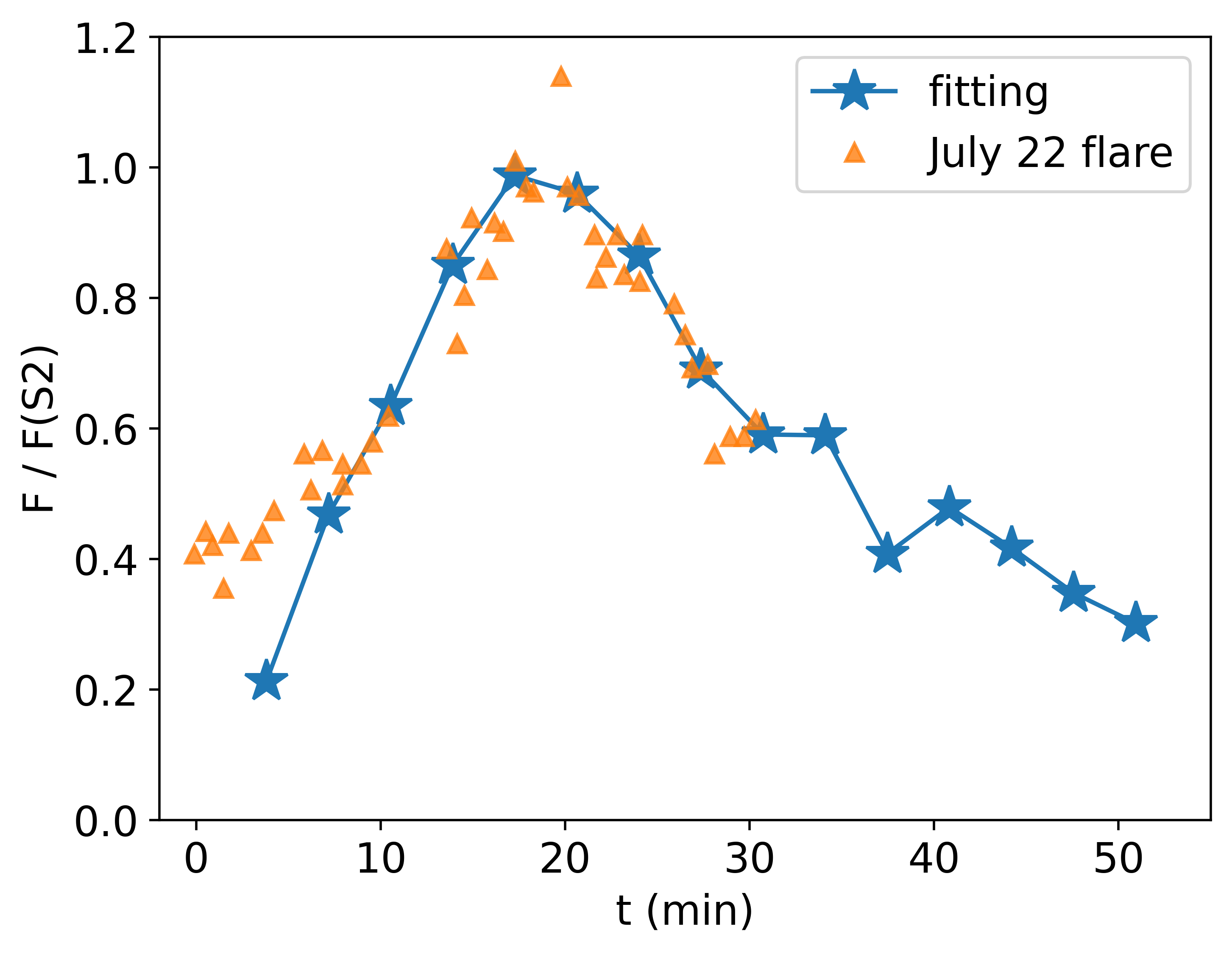}
    \caption{Variability predicted by our model (denoted by the blue stars) and its  comparison with the observational data of the July 22 flare (denoted by the orange triangles). The latter is taken from Fig. 2 of \citet{GRAVITY2018}. }
    \label{fig:flux}
\end{figure}

In Figure \ref{fig:flux}, we show the predicted light curve and its comparison with the observed light curve of the July 22 flares. In our model, the variability is because of the evolution of the nonthermal electrons, which is calculated by solving equation (\ref{nonthermale}), and the change of the magnetic field surrounding the flux rope when it moves outward. As we can see from the figure, our model nicely reproduce the observed light curve, including the normalization of the flux and the profile. Given that there are no free parameters to adjust in our model when fitting the light curve, such an agreement is exciting. 

\begin{figure}
	\includegraphics[width=\columnwidth]{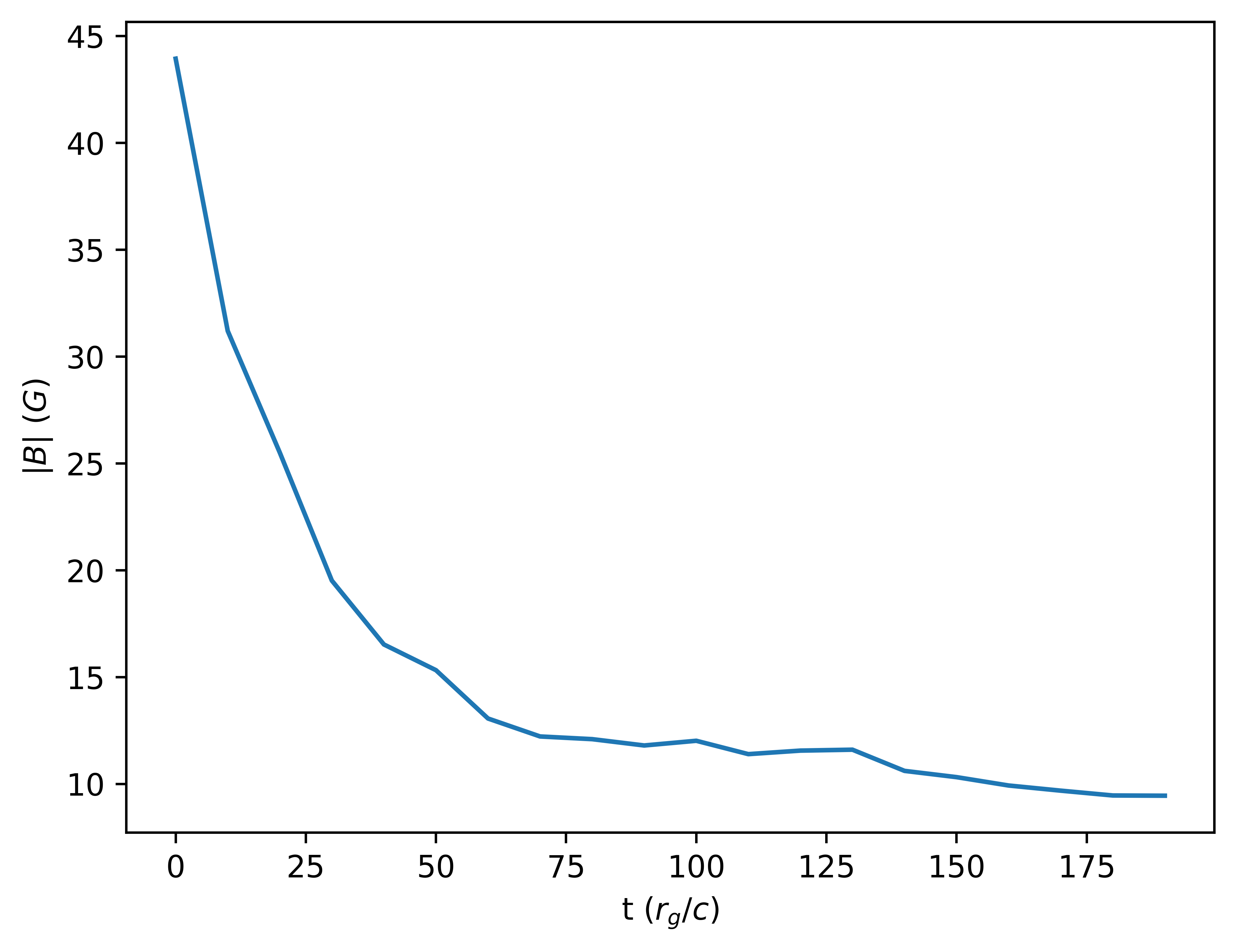}
    \caption{Time evolution of the magnetic field magnitude in the flux rope center. The field magnitude monotonically decreases as the flux rope is ejected away from the black hole.}
    \label{fig:Bfield}
\end{figure}

\begin{figure}
	\includegraphics[width=\columnwidth]{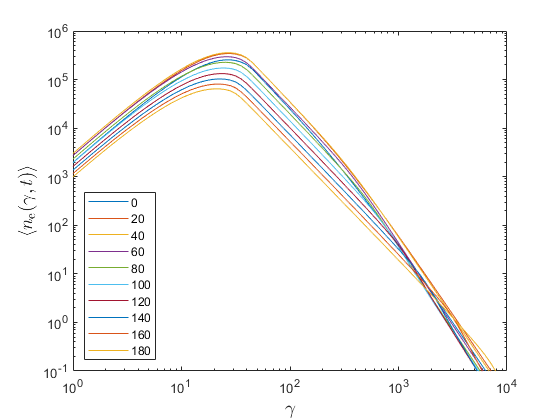}
    \caption{Time evolution of the mean number density of the non-thermal electrons in the flux rope. Colors mark the evolution time ranging from 0 to 180 $r_g/c$. }
    \label{fig:nth}
\end{figure}

To analyze the underlying reason for the rise and decay of the light curve, Figure \ref{fig:Bfield} shows the evolution of the magnetic field in the flux rope, while  the evolution of the number density of nonthermal electrons is shown in Figure \ref{fig:nth}. We can understand from the figures that the rise of the light-curve is mainly caused by the injection of the non-thermal electrons during magnetic reconnection. The decay part, however, is likely caused by multiple factors. On the one hand, as the flux rope moves away from the black hole, magnetic field strength around the rope, which directly determines the synchrotron emissivity, decreases almost monotonically (Figure \ref{fig:Bfield}). In addition, the magnetic field around the reconnection inflow region underneath the flux rope, which scales the injection term, also declines. 

\subsection{Polarization}
\label{polarization}
The most serious problem in our previous work of \citet{Lin2023} is that the predicted polarization is not consistent with observations. Specifically, the polarization angle is predicted to rotate twice during one orbital period, while observations show only once. We have speculated in that work that, this problem is most likely because of the unrealistic magnetic field configuration adopted in the model. In that work,  an unrealistic magnetic field configuration has been adopted in order to obtain an analytical solution. 

\begin{figure*}
	\includegraphics[width=1.\textwidth]{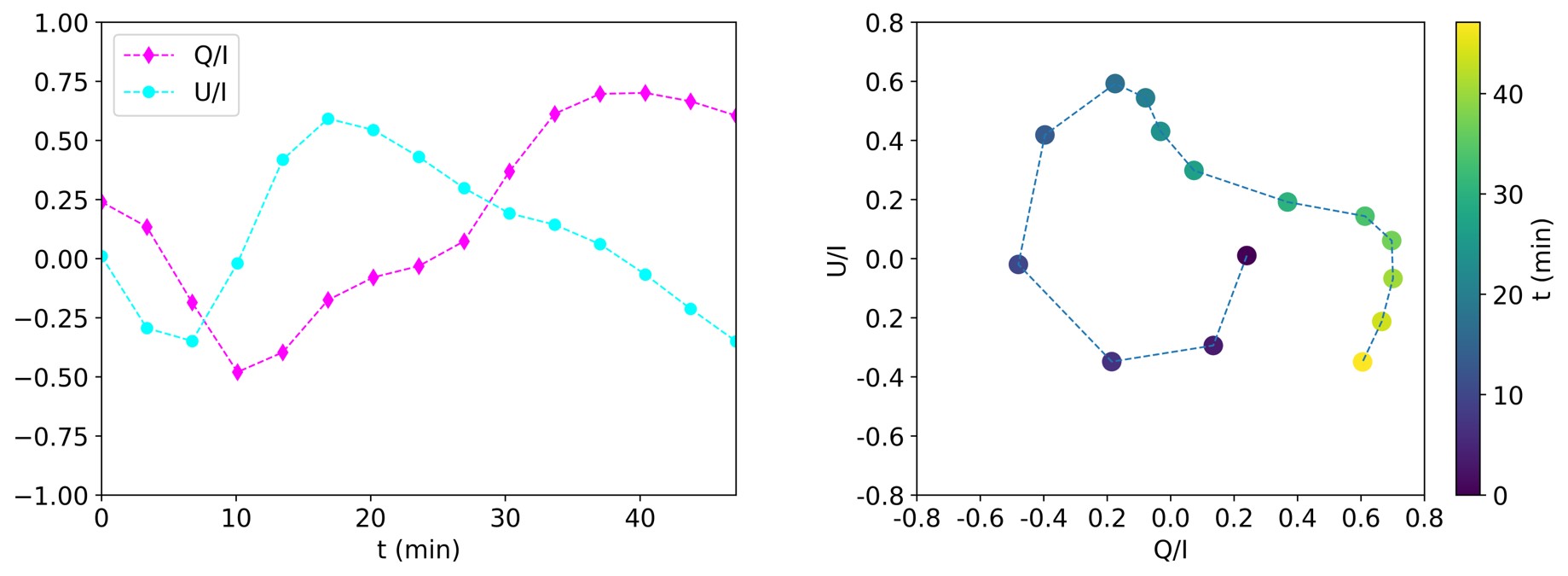}
    \caption{Stokes parameters $Q/I$ and $U/I$ during the flare. The polarization angle shows a continuous rotation in the $Q-U$ map, with a period comparable to the orbital period. } 
    \label{fig:polarization}
\end{figure*}

In the present work, we have realistic magnetic field configuration taken from our GRMHD numerical simulations thus we expect we can obtain polarization result consistent with observation. Figure \ref{fig:polarization} shows the calculation results.  We do find from the figure that the rotation in the polarization $Q-U$ map now has a period of polarization and orbit consistent with observations. The ordered and dominant poloidal magnetic field in MADs is crucial to explain the continuous rotation of polarization angle. However, since the magnetic field is not fully uniform and also evolves with time, the loop in the $Q-U$ map thus is not completely closed and more irregular in shape compared with analytic models \citep{Vos2022}. 

The mean degree of polarization is $\sim 50\%$, higher than that given by GRAVITY, which is $\sim 20-40\%$. We think that this is at least partly because we use a simple spherical-like hot spot to characterize the flux rope. In realistic situation, the radiating flux rope is stretched in the azimuthal direction and form a filament-like structure. Different parts of the flux rope will contribute to the observed radiation. Since the magnetic field is not spatially uniform, the  polarization degree will become lower when the radiation from various part of the flux rope is summed up.

\subsection{Physical mechanism of the periodic formation and ejection of flux ropes}
In the cases of both SANE and MAD, it is found that the reconnection and the formation of flux rope seem to occur quasi-periodically, with the period being $\sim 1000 r_g/c$. This value roughly corresponds to the rotation timescale at $20\sim30 r_g$ \citep{Miki2022}. This periodicity was speculated in \citet{Miki2022} to be because that the differential rotation of the accretion flow drives the magnetic reconnection. The detail of this process is still unclear. But we speculate that after a time interval of about one orbital time, the magnetic field lines are twisted enough so that the helicity of the magnetic field lines becomes high enough thus reconnection can occur \citep{2005ARA&A..43..103Z}. It is interesting to note that, in the case of MAD, the occurrence of magnetic reconnection and the subsequent formation of flux rope have more dramatic effect compared to the case of SANE. In this case, the reconnection and formation of flux rope seem to be accompanied by the ejection of  large amount of magnetic flux from the region close to the black hole, i.e., a magnetic eruption event. This seems to be supported by the variability of the magnetic flux shown in Figure \ref{fig:mdot}, from which the same period is present. 
The eruption interval in our simulation is $\sim 1000 r_g/c$ which corresponds to $\sim 5.5$ hours in the case of Sgr A*, consistent with the timescale of its quiescent state. Other possible applications of the periodicity of the formation and ejection of flux ropes have been discussed in \citet{Miki2022}.


\section{Summary}\label{sec:discussion}
\label{sec:6}
High-resolution GRAVITY observations of  the infrared flares of Sgr A* have provided unprecedented abundant information to constraint its nature. The observation results suggest a scenario in which during the flare a hot spot rotates with the black hole with a rotation radius around $\sim 10 r_{\mathrm{g}}$, with the value increasing with time. Moreover, the combination of the radius and rotation time indicates that the hot spot seems to rotate with a super-Keplerian velocity. In \citet{Yuan2009}, we have proposed a scenario to explain the physical mechanism of the flares and the associated formation of an ejected flux rope which is responsible for the  observed hot spot. In this scenario, magnetic reconnection occurs due to turbulent motion and the differential rotation of the accretion flow. The reconnection, on the one hand, accelerates electrons and produce flares; on the other hand, results in the formation of the flux rope, which is ejected out by Lorentz force.  \citet{Li2017} and \citet{Lin2023} have calculated the predicted radiation of the process and applied to the interpretation of the flares in Sgr A*.

However, both \citet{Li2017} and \citet{Lin2023} are analytical works and some simplifications have been adopted in their calculations.  In the present work, we revisit this problem by directly using the GRMHD numerical simulation data. 
The simulation data of both SANE and MAD have been analyzed.  We have first studied the distribution and structure of the current sheets, which are indicators of reconnection. We found current sheets in the SANE simulation are more curly and widespread in the accretion flow, while those in the MAD are flat and concentrated near the equatorial plane and jet sheath region. Such a different distribution is because the magnetic field in MAD is much stronger thus the accretion flow is  less turbulent. We find that  the flux ropes formed in the MAD is more suitable to interpret the flares of Sgr A*. The presence of the strong current sheet in the equatorial plane is due to the accumulation of magnetic flux and the resulting strong magnetic pressure. We find from the data that magnetic reconnection occurs at the place of current sheet and flux rope are formed consequently. 

 The flux rope, as the product of reconnected magnetic field lines, encloses the accelerated electrons and is responsible for the observed radiation.  We find that the flux ropes formed in the jet region are prone to be ejected out helically while those formed near the equatorial plane will eventually be accreted into the black hole. We chose a fiducial flux rope initially born in the inner polar corona region $\sim 8 r_{\mathrm{g}}$ away from the black hole and traced its trajectory in the following $200r_{\mathrm{g}}/c$. 
  To account for the radiation from the flux rope, we have calculated the injection rate of the non-thermal electrons due to reconnection as well as their energy distribution. Simplifications have to be made in these calculations, which is the major weakness and limitation of the present work. 
 We assume that the accelerated electrons in the reconnection  will not stay in the current sheet but will be advected with the reconnection outflow. These electrons are finally accumulated in the flux rope. The time-dependent energy distribution of the electrons has been solved by considering the injection due to acceleration, synchrotron radiation, and adiabatic expansion. 
We have calculated the radiation from the flux rope using ray-tracing approach. The radiating flux rope exhibits in the form of a hot spot. Its  trajectory has been obtained and compared to the GRAVITY observations. Good consistency is found, including the increasing radius of the observed hot spot to the black hole. The rotation speed of the hot spot is somewhat slower than observations, but this is reconciled once the relativistic poloidal motion of the flux rope is properly considered.
We have also calculated the time-dependent radiation of the flux rope and compared with the observed light curve. Good consistency is again found. The rising part of the lightcurve is mainly because of the injection of nonthermal electrons during the reconnection acceleration. The delay part of the light curve are mainly because of the decrease of the magnetic field around and beneath the flux rope, which determines the radiative cooling and the injection of energetic electrons, respectively. The predicted polarization is also calculated and compared with that obtained by GRAVITY. One problem of \citet{Lin2023} is that the predicted polarization period is not consistent with observations.  By using a more realistic magnetic field, now we can successfully explain the observed polarization period.  We argue that the ordered poloidal dominant magnetic field in MADs plays an essential role in reproducing the characteristic loops in the $Q-U$ map.

\section*{Acknowledgements}
We thank the referee for the constructive comments and suggestions. We thank YP Li for insightful discussions. This work is supported in part by the Natural Science Foundation of China (grants 12133008, 12192220, and 12192223).  The calculations have made use of the High Performance Computing
Resource in the Core Facility for Advanced Research Computing
at Shanghai Astronomical Observatory. 

\section*{Data availability}
The data underlying this article will be shared on reasonable request to the corresponding author.

\bibliographystyle{mnras}
\bibliography{lx.bib}{}

\appendix
\section{Resolution impact on the accretion rate and magnetic flux}
\label{appendix:A}

\begin{figure*}
	\includegraphics[width=\columnwidth]{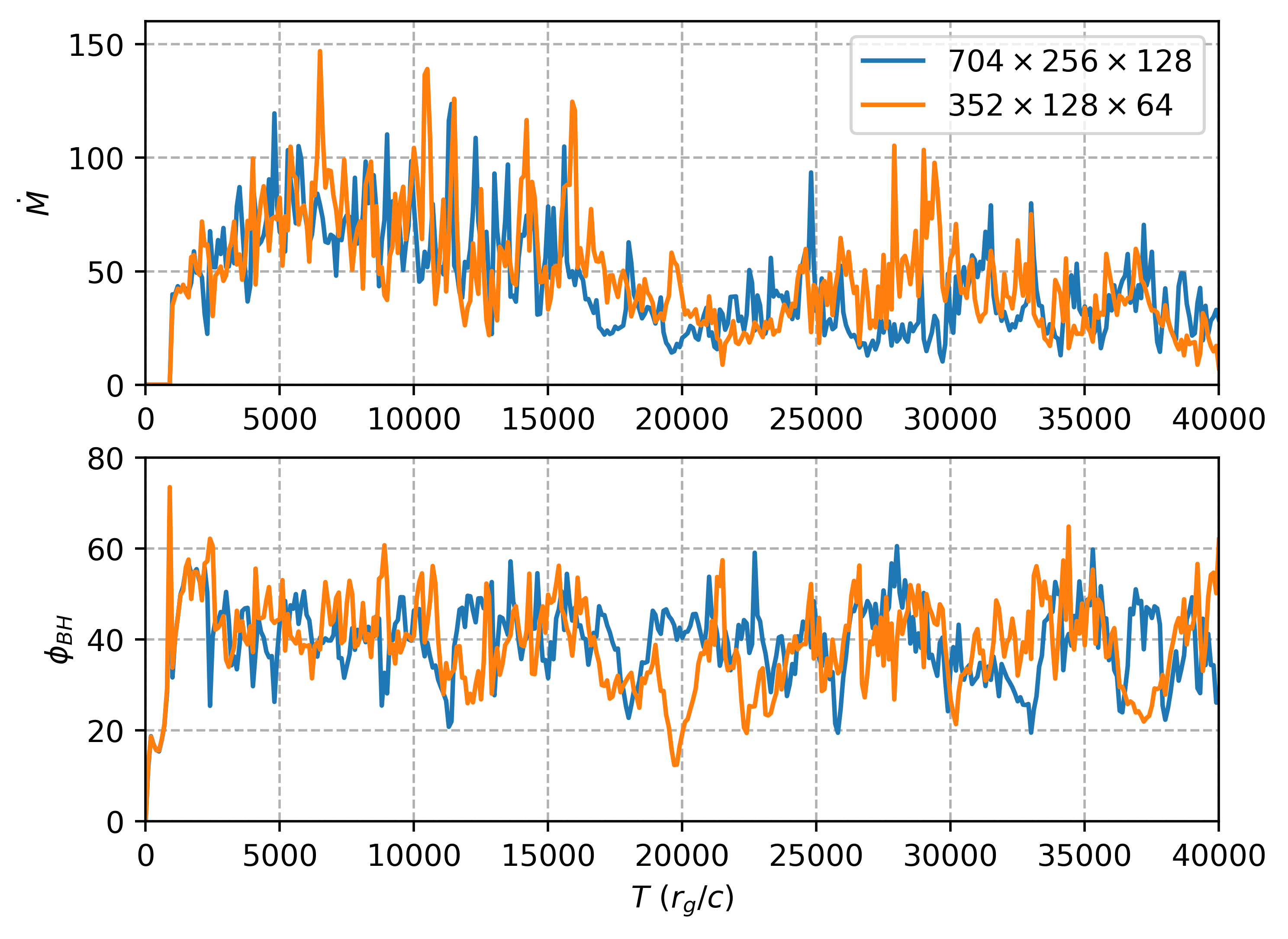}
    \caption{Mass accretion rate and magnetic flux of MAD98 with different resolution.}
    \label{fig:resolution}
\end{figure*}

\begin{figure*}	\includegraphics[width=1.\textwidth]{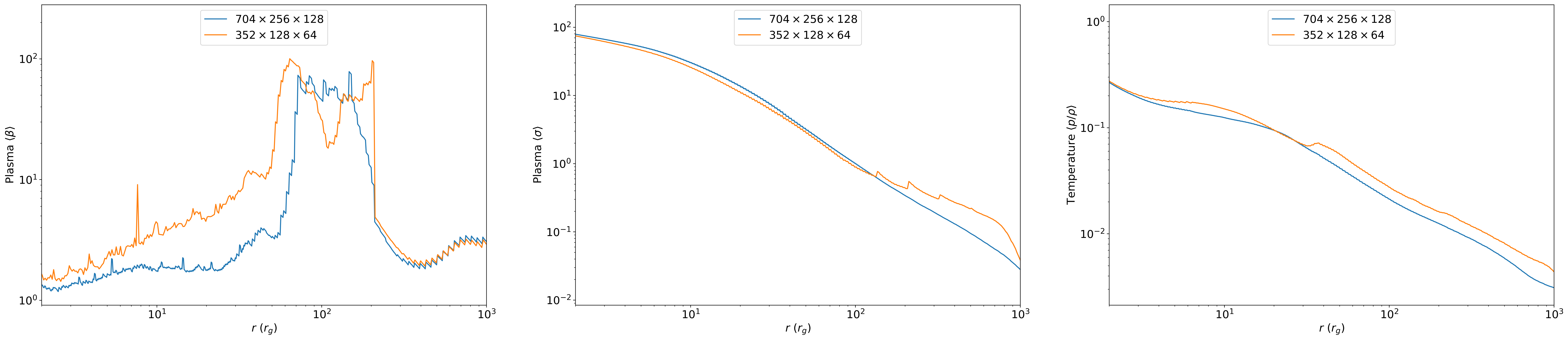}
    \caption{Mean Shell-averaged plasma $\beta$, $\sigma$ and temperature $p/\rho$ profiles as functions of radius $r$ averaged over the time interval $T=20000-30000$ for the two MAD98 models with a different resolution.}    \label{fig:shell}
\end{figure*}

To study the resolution convergence, we compared our present high-resolution simulation data with previous low-resolution data. The accretion rate and magnetic flux of MAD98 are calculated for a representation. As is shown in Figure \ref{fig:resolution}, both quantities from different resolutions have similar evolution with time. The saturation limit for MAD98 is also consistent under two resolution simulations. Figure \ref{fig:shell} shows the shell-averaged quantities after being time averaged. Except the shell-averaged plasma $\beta$ in low resolution shows a higher value at $r<50$, others quantities are generally matched. 

\bsp	
\label{lastpage}

\end{document}